\documentclass[a4paper]{article}
\usepackage[margin=25mm]{geometry}
\usepackage{amsmath}
\usepackage{amsfonts}
\usepackage{amssymb}
\usepackage{graphicx}
\usepackage{verbatim}
\usepackage{authblk}
\usepackage{cite}
\usepackage{amsthm}
\usepackage[frozencache,cachedir=minted-cache]{minted}

\usepackage{lineno,hyperref}
\usepackage{makecell}
\usepackage{amsmath}
\usepackage{comment}
\usepackage{mathtools}
\usepackage{color}
\usepackage[normalem]{ulem}
\hypersetup{colorlinks=true,citecolor=blue}
\let\vec\mathbf
\usepackage{courier}

\usepackage[title]{appendix}

\newcommand{\vardbtilde}[1]{\tilde{\raisebox{0pt}[0.85\height]{$\tilde{#1}$}}}

\title{Explicit energy-conserving modification\\ of relativistic PIC method}

\author{\small Arkady~Gonoskov}

\affil{\small Department of Physics, University of Gothenburg, SE-41296 Gothenburg, Sweden}

\date{} 

\begin{document}
\maketitle

\begin{abstract}
The use of explicit particle-in-cell (PIC) method for relativistic plasma simulations is restricted by numerical heating and instabilities that may significantly constrain the choice of time and space steps. To {partially} eliminate these limitations we consider a possibility to enforce exact energy conservation by altering the standard time step splitting. {Instead of updating particles in a given field and then the field using the current they produce, we consider subsystems that describe the coupling of a single particle and the field at the nearby nodes and solve them with enforced energy conservation sequentially for all particles, which is completed by the field update with zero current.} Such {an approach} is compatible with various advances, {ranging from accounting for additional physical effects to the use of numerical-dispersion-free field solvers, high-order weighting shapes and particle push subcycling.} To facilitate further considerations and use, we provide a basic implementation in a 3D, relativistic, spectral code $\pi$-PIC, which we make publicly available. {The method and its implementations are verified using simulations of cold plasma oscillations, Landau damping and relativistic two-stream instability. The capabilities of the method to deal with large time and space steps are demonstrated in the problem of plasma heating by intense incident radiation.}
\end{abstract} \hspace{10pt}

\section{Introduction}
The particle-in-cell (PIC) method \cite{birdsall.1984} has a long history of developments that gradually made it suitable for a vast range of numerical studies in fundamental and applied plasma physics. Some of the developments are related to efficient use of computational resources and the possibility to perform computations in parallel on modern supercomputers \cite{fonseca.lncs.2002, bastrakov.jcs.2012, bussmann.ichpc.2013, derouillat.cpc.2018, myers.pc.2021, bird.tpds.2022, fedeli.cs.2022}. Other developments concern the inclusion of physical processes, such as ionization or the effects due to strong-field quantum electrodynamics \cite{duclous.ppcf.2011, nerush.prl.2011, elkina.prstab.2011, sokolov.pop.2011, ridgers.jcp.2014, gonoskov.pre.2015, arber.ppcf.2015, lobet.jpcs.2016}, as well as the solution of the concomitant algorithmic difficulties, such as particle ensembles resampling \cite{rjasanow.jcp.1996, rjasanow.jcp.1998, lapenta.jcp.1994, lapenta.cpc.1995, assous.jcp.2003, welch.jcp.2007, timokhin.mnras.2010, vranic.cpc.2015, pfeiffer.cpc.2015, martin.jcp.2016, luu.cpc.2016, faghihi.jcp.2020, muraviev.cpc.2021, gonoskov.cpc.2022} to compensate for the growth of particle number due to ionization or particle/photon production. Finally there are developments related to the elimination of fundamental limitations, which can potentially facilitate a broad range of PIC method applications.  

One such limitation, which remains a long-standing problem, is the numerical dispersion of field solvers. Although dispersion-free spectral solvers have been known from late 70's \cite{haber.cnsp.1973, buneman.jcp.1980}, their practical use is limited by difficulties of arranging parallel computations. That is why parallel codes are commonly based on grid-based field solvers (exceptions include codes described in Refs.~\cite{vay.jcp.2013, gonoskov.phd.2013, fedeli.cs.2022}). One way to overcome the limitation while still using grid-based methods is to modify {the finite-difference time-domain (FDTD)} scheme \cite{yee.tap.1966} so that the numerical dispersion is suppressed/eliminated along the directions of interest \cite{pukhov.jpp.1999, cowan.prstab.2013, lehe.prstab.2013, blinne.cpc.2018, pukhov.jcp.2020}.

One other long-standing problem of the PIC method is the lack of exact energy conservation, which leads to numerical heating and instabilities \cite{birdsall.1984}, thereby restricting the choice of time and space steps. The early developments related to energy-conserving PIC methods were carried out in early 70's by Lewis \cite{lewis.jcp.1970} and Langdon \cite{langdon.jcp.1973}. The idea of energy-conserving simulation is that the state of particles and the sate of electromagnetic field have to be advanced in time concurrently so that the energy conservation is enforced. This can be done by implicit scheme derived from Lagrangian formulation. Nevertheless this leads to the necessity of solving nonlinear system of equations, those number grows with the number of particles. Using Newton Krylov solver makes it possible to exactly preserve energy, under assumption of convergence. This approach was developed and used for the implementation of implicit PIC solvers in Refs.~\cite{kim.dpp.2005, markidis.phd.2010, markidis.jcp.2011, chen.jcp.2011, taitano.jsc.2013}. {In 2020 Chen, Chacon, and their collaborators proposed an energy- and charge-conserving semi-implicit algorithm that uses multiple implicit iterations for each particle to iteratively solve a system that describes energy coupling between particles and fields (the focus on this coupling is also fundamental to the method we develop here) \cite{chen.jcp.2020}.} Some recent developments on relativistic energy-conserving implicit PIC method are reported in Ref.~\cite{li.arxiv.2022}. 

In 2017 Lapenta proposed a semi-implicit method (referred to as Energy Conserving Semi-Implicit Method (ECSIM)) that achieves the property of exact energy conservation {by solving a linear system with the size proportional to the number of grid nodes, rather than presumably much larger number of particles as in fully implicit methods} \cite{lapenta.jcp.2017}. The idea of Lapenta method is to include the linear response of particles to the field into Maxwell’s equations solved using $\theta$-scheme \cite{brackbill.jcp.1982, vu.cpc.1992} and advance particles using explicit for coordinate and implicit for velocity leap-frog scheme with electric field at the $\theta$ time level \cite{langdon.jcp.1983, hewett.jcp.1987}. Although in Ref.~\cite{lapenta.jcp.2017} the method was developed for non-relativistic case, it is possible that the approach can be generalized to relativistic case and be useful even in highly non-linear simulations. 

Nevertheless, for many studies that concern highly relativistic non-linear plasma dynamics, it is common to use explicit PIC method with sufficiently small time and space steps to avoid numerical heating and instabilities (these effects can be also suppressed by high-frequency field filtering). The choice of explicit methods can be related to the readiness for parallel implementation and overall possibility to use smaller time and space steps than that possible with implicit methods under similar computational demands. 

In this paper we describe a possibility to modify the standard relativistic PIC scheme so that we enforce exact energy conservation while still keeping the scheme explicit and permitting the use of dispersion-free spectral field solver. This is achieved by splitting the time step into two actions: current-free energy-conserving advancement of the field and energy-conserving local coupling of the field (current-related part) with the dynamics of each particle. The coupling is performed sequentially for all particles{, which can be seen as a multiple application of the Strang splitting \cite{strang.jna.1968, mclachlan.an.2002}. Although this significantly restricts opportunities for parallel processing, there is still some possibility for parallel implementation, which we describe in Sec.~\ref{sec_implementation} and use in $\pi$-PIC code.} We discuss options for achieving the desired coupling and describe a basic option of corresponding particle-field pusher. In the limit of small time step the scheme tends to the update of particles' momenta and the current-related change of the field prescribed by the conventional PIC method. However, by performing the particle momentum update together with the update of the electric field at the nearby nodes, we ensure that the total energy remains unchanged to within machine accuracy. {We should note that the exact conservation of energy can prevent numerical heating but not necessarily all numerical instabilities, for example finite-grid (or aliasing) instabilities for drifting plasmas (see details in Ref.~\cite{barnes.cpc.2021} and in Chapter 8 in the book by Birdsall and Langdon \cite{birdsall.1984}). In summary the proposed method has the following properties}:
\begin{itemize}
    \item the method preserves energy even in case of highly nonlinear, relativistic dynamics and thus is not subject to numerical {heating};
    \item the method doesn't include implicit parts and admits parallel implementation;
    \item the method is compatible with an arbitrary field solver, including the spectral solver, which is free from numerical dispersion;
    {\item the method admits an implementation that has the second order accuracy in time.}
\end{itemize}
We conjecture that due to {these} properties the approach described in this paper might be found useful in various numerical studies of relativistic, as well as non-relativistic plasma dynamics.

The paper is arranged as follows. In Sec.~\ref{sec_modified_PIC} we motivate and describe the modified PIC method. In Sec.~\ref{sec_coupling} we derive one basic implementation of the energy-conserving particle-field pusher used in the method and discuss possible improvements. The use of the standard spectral solver is detailed in Sec.~\ref{sec_spectral_solvers} for completeness. {In Sec.~\ref{sec_acc_improvements} we analyse the accuracy order, develop the variant with the second order accuracy and discuss other possible improvements. In Sec.~\ref{sec_implementation} the basic implementations of the method included in $\pi$-PIC are described and tested using the simulation of Landau damping and relativistic two-stream instability.} We study basic properties of the method using plasma oscillations in Sec.~\ref{sec_properties}. In Sec.~\ref{sec_verification} we present a more sophisticated {verification} test and show the capabilities of the proposed method. We summarise the work in Sec.~\ref{sec_conclusion}.  

\section{Modified PIC scheme}
\label{sec_modified_PIC}

When particle dynamics becomes relativistic the idea of enforcing energy conservation faces several difficulties. The first difficulty is that the velocity response becomes nonlinear due to Lorentz factor. One way to combat this is to linearize the response near the current state. Nevertheless, it is possible that the change of particle momentum and velocity over a single time step is not small enough. Recent studies showed \cite{tangtartharakul.jcp.2021} that in some cases of interest strong magnetic fields in combination with low particle energies cause difficult-to-resolve gyration, thus requiring advanced particle pushers \cite{vay.pop.2008} and subcycling \cite{arefiev.pop.2015}. Although this aspect is clearly important for ensuring accurate numerical results, it doesn't seem the primary difficulty for the energy conservation because even standard Boris pusher \cite{boris.p.1970} preserves energy when accounting {only} for the magnetic field.

The second difficulty is that over a single time step $\Delta t$ a relativistic particle can traverse a distance comparable with the space step $\Delta x$. (If the spectral field solver is used, the time step can be equal or even larger than $\Delta x/c$, where $c$ is the speed of light.) In case of significant change of particle velocity over $\Delta t$, the coordinate becomes another nonlinear part in the response to the field (along with the velocity itself). Moreover, the update of coordinate by an explicit method based on "old" velocity (as Lapenta method prescribes) unambiguously determines the current, depriving us of the freedom critical for enforcing energy conservation. This is because the current has to be consistent with the charge relocation to comply with the current continuity equation. This consistency is known to be important in relativistic simulations and is perfectly addressed by Esirkepov scheme for current deposition \cite{esirkepov.cpc.2001}. (For spectral solvers, there are also options of corrections via direct solution of the Poisson's equation.)

To identify requirements and possibilities for enforcing energy conservation it is instructive to consider 1D problem of electric field interacting with a moving charged particle. Within a sufficiently large time step the energy conservation can become intricate in two different ways: (1) the particle is being accelerated and this process halts when the field energy is exhausted and (2) the field is being set up by the moving particle and this totally exhausts its energy turning the particle in the opposite direction. If both happens within a single step, an oscillating behaviour occurs. This consideration indicates one notable point: to achieve energy-conservation it is not possible to advance particle location based on the "old" momentum because we do not know in advance whether the particle will propagate that far. That is why we choose to reverse the common logic: we first determine current that is consistent with momentum change and then determine the next particle location to satisfy the charge continuity equation. 

To implement the outlined principle we start from the standard formulation of coupled evolution of classical electromagnetic field and charged particles (CGS units are used):
\begin{align}
    &\frac{\partial}{\partial t} \vec{B} = -c \nabla \times \vec{E},\\
    &\frac{\partial}{\partial t} \vec{E} = c \nabla \times \vec{B} - 4\pi \vec{J},\\
    &\vec{J} = \sum_{i=1}^N q_i \vec{v}_i w \left(\vec{r}_i - 
    \vec{r}\right),\\
    &\frac{\partial}{\partial t} \vec{r}_i = \vec{v}_i = c\vec{p}_i \left(1 + p_i^2\right)^{-1/2}, \ i = 1, ..., N,\\
    &\frac{\partial}{\partial t} \vec{p}_i = \frac{q_i}{m_i c} \left(\vec{E}\left(\vec{r}_i\right) + \frac{1}{c} \vec{v}_i \times \vec{B}\left(\vec{r}_i\right)\right), \ i = 1, ..., N,
\end{align}
where $\vec{B}$, $\vec{E}$ are magnetic and electric field vectors; $\vec{J}$ is the current; $\vec{r}_i$, $\vec{p}_i$, $\vec{v}_i$, $q_i$ and $m_i$ are the $i$-th particle coordinate, momentum (given in units of $m_i c$), velocity, charge and mass, respectively; $N$ is the number of particles; $w(\cdot)$ quantifies particles' form factor. To enforce energy conservation we consider time step splitting into steps that advance the state according to subsystems of our choice. We first notice that the field evolution without current do not directly concern the energy coupling between particles and fields. This defines a useful choice for the first subsystem:
\begin{align}
    &\frac{\partial}{\partial t} \vec{B} = -c \nabla \times \vec{E},\\
    &\frac{\partial}{\partial t} \vec{E} = c \nabla \times \vec{B},
\end{align}
which can be exactly {(to within errors related to the limited number of Fourier modes for the discrete grid)} solved using spectral solver to advance the state of field with exact energy conservation (details are given in Sec.~\ref{sec_spectral_solvers}). Since magnetic field does no work, the effect of magnetic field on momentum $\vec{p}_i$ is another part that doesn't concern energy exchange and defines a useful set of independent subsystems:
\begin{equation}
    \frac{\partial}{\partial t} \vec{p}_i = \frac{q_i}{m_i c} \left(1 + p_i^2\right)^{-1/2} \vec{p}_i \times \vec{B}\left(\vec{r}_i\right), \ i = 1, ..., N.
    \label{subsys_b}
\end{equation}
Each subsystem (\ref{subsys_b}) can also be solved exactly by rotating $\vec{p}_i$ about $\vec{B}\left(\vec{r}_i\right)$, or approximately, using Boris scheme, which also preserves energy.

The remaining parts account for the current in Maxwell's equations and for the electric part of the Lorentz force in the equation of motion. These parts are responsible for the energy coupling between the field and particles. We now notice that we can split the current into contributions of each particle $\vec{J}_i$ and separate the remaining subsystem into $N$ subsystems. The $i$-th subsystem describes local coupling between the field and $i$-th particle:   
\begin{align}
    &\frac{\partial}{\partial t} \vec{p}_i = \frac{q_i}{m_i c} \vec{E}\left(\vec{r}_i\right), \label{subsys_c1}\\
    &\frac{\partial}{\partial t} \vec{E} = - 4\pi \vec{J}_i,\label{subsys_C2}\\
    &\vec{J}_i = q_i c\vec{p}_i \left(1 + p_i^2\right)^{-1/2} w \left(\vec{r}_i - 
    \vec{r}\right),\label{subsys_c3}\\
    &\frac{\partial}{\partial t} \vec{r}_i = c\vec{p}_i \left(1 + p_i^2\right)^{-1/2}\label{subsys_c4}
\end{align}
Even thought the particles can overlap we intentionally split the coupling between the field and particles into $N$ independent subsystems, which we choose to advance sequentially to be able to enforce energy conservation for each subsystem. This corresponds to the principle of particle-mesh approach that implies that particles do not interact with each other directly within a single time step, but instead interact through the field defined on a mesh/grid. In what follows we omit index $i$ because the subsystems for each particle are independent and their consideration is identical.

To construct an energy-conserving numerical method for the outlined step splitting we need to formulate a consistent form of subsystem \eqref{subsys_c1} -- \eqref{subsys_c4} for the field defined on a grid. Although a complete treatment with advanced approaches can be useful here, we are primarily aiming at preserving energy and thus use the following simplification. We assume that the interaction with the nearest grid values of the field is represented by coefficients $c_j > 0$ that do not change in time during a single time step:
\begin{equation}
    \forall \ \tau \in [t, t + \Delta t]: \ \vec{E}(\vec{r}(\tau), \tau) = \sum_{j=1}^{M}c_j\vec{E}_j(\tau),
\end{equation}
where index $j$ enumerates the nearest $M$ grid nodes involved in the interaction and $\vec{E}_j(\tau)$ is the electric field vector at the $j$-th node. Certainly, the normalization condition must be satisfied:
\begin{equation}
    \sum_{j=1}^M c_j = 1.
    \label{c_norm}
\end{equation}
The coefficients are defined by the particle form factor and can be determined in different ways depending on necessary accuracy. In general we need to compute a cumulative overlap. This can be done assuming that the particle propagates with the "old" velocity:
\begin{equation}
    c_j = c_j(\vec{r}(t), \vec{p}(t)) = {\Delta t^{-1}} \int_{\tau = 0}^{\Delta t} d \tau \iiint_{V_g} d^3r_g S\left(\vec{r}(t) + \tau \frac{c \vec{p}(t)}{\sqrt{1+p^2(t)}} - \vec{r}^{(j)} -\vec{r}_g\right),
\end{equation}
where $S(\vec{r})$ describes the shape of the particle, $V_g$ denotes the volume of each cell centered at the $j$-th node $\vec{r}^{(j)}$. {To comply with eq.~\eqref{c_norm} we require that function $S(\vec{r})$ is normalized, i.e. $\iiint d^3r S(\vec{r}) = 1$.} If the integration can be done analytically the procedure becomes computationally cheap. Alternatively one can approximate the value of the integral using one or several points within the time interval. One can further increase accuracy of the final particle location at the end of the time step by running a predictor-corrector cycle, using the entire procedure described below. Nevertheless, it is not clear whether such more accurate calculations with non-propagating field within one time-step would me more efficient than the overall reduction of the time step for the simplest method. Since our primary goal is to consider energy conservation, in the provided implementation we use the simplest option of integral estimation based on the middle point and cloud-in-cell (CIC) weighting for the shape function.

{The method is applicable for arbitrary weighting scheme. To exemplify the use of the method let us detail the case of CIC weighting, which we use in the provided implementation. To do so we define the CIC weighting function:
\begin{equation}
    W_{CIC}(\vec{r}) = T\left(x/\Delta x\right)
    T\left(y/\Delta y\right)
    T\left(z/\Delta z\right),
\end{equation}
where $x$, $y$, $z$ are the Cartesian components of $\vec{r}$; $\Delta x$, $\Delta y$, $\Delta z$ are the corresponding space steps ($V_g = \Delta x \Delta y \Delta z$), and $T(\cdot)$ is the triangular shape function:
\begin{equation}
    T(u) = \begin{cases}
        1 - \left|u\right|, \left|u\right| < 1,\\
        0, \ \left|u\right| \geq 1.
    \end{cases}
\end{equation}}

Assuming that coefficients $c_j$ are fixed for the duration of time step, we define a grid-base version of subsystem \eqref{subsys_c1} -- \eqref{subsys_c3}:
\begin{align}
    &\frac{\partial}{\partial t} \vec{p} = \frac{q}{m c} \sum_{j = 1}^M c_j \vec{E}_j, \label{subsys_c1n} \\
    &\frac{\partial}{\partial t} \vec{E}_j = -4\pi c_j \bar{\vec{J}}, \ j = 1, ..., M, \label{subsys_c2n}\\
    &\bar{\vec{J}} = \frac{q c}{V_g}\frac{\vec{p}} {\sqrt{1 + p^2}}, \label{subsys_c3n}
\end{align}
where $\bar{\vec{J}}$ has the meaning of the current produced by the particle. To demonstrate that this formulation is consistent in terms of energy conservation, let us explicitly show that the energy is preserved. To do so we multiply both sides of \eqref{subsys_c2n} on $\vec{E}_j$, sum over all $j = 1, ..., M$, substitute \eqref{subsys_c3n} and \eqref{subsys_c1n}, and finally obtain
\begin{equation}
    \sum_{j=1}^M \vec{E}_j \cdot \frac{\partial}{\partial t} \vec{E}_j = -4\pi \frac{m c^2}{V_g} \frac{\vec{p}}{\sqrt{1 + p^2}} \cdot \frac{\partial \vec{p}}{\partial t}.
\end{equation}
Multiplying both sides on $V_g/4\pi$, we finally get
\begin{equation}
    \frac{\partial}{\partial t}\left(\frac{V_g}{8\pi}\sum_{j=1}^M E_j^2 + m c^2 \sqrt{1 + p^2}\right) = 0,
    \label{eq_en_cons}
\end{equation}
which explicitly shows that the system \eqref{subsys_c1n} -- \eqref{subsys_c3n} preserves total energy of the particle and the field defined on a grid. This means that if we advance the state according \eqref{subsys_c1n} -- \eqref{subsys_c3n} so that the energy is preserved, then the whole method preserves energy.

Now let us show that eqs.~\eqref{subsys_c1n} -- \eqref{subsys_c3n} {correspond to a well-posed problem, i.e. define a system that has} a single solution for arbitrary initial conditions. To show this we substitute \eqref{subsys_c3n} into \eqref{subsys_c2n}, multiply both sides by $c_j$ and sum over all $j = 1, ..., M$:
\begin{equation}
    \sum_{j=1}^M c_j \frac{\partial \vec{E}_j}{\partial t} = -4\pi \frac{qc}{V_g} \left( \sum_{j=1}^M c_j^2 \right) \frac{\vec{p}} {\sqrt{1 + p^2}} = -4\pi \frac{qc}{V_g} \xi \frac{\vec{p}} {\sqrt{1 + p^2}}, \label{eq_cdEjdt}
\end{equation}
where we defined a constant
\begin{equation}
    \xi = \sum_{j=1}^M c_j^2 > 0.
\end{equation}
We then take time derivative of \eqref{subsys_c1n} and substitute \eqref{eq_cdEjdt} into the right hand side:
\begin{equation}
    \frac{\partial^2}{\partial t^2} \vec{p} = - 4 \pi \frac{q^2}{m V_g} \xi \frac{\vec{p}} {\sqrt{1 + p^2}}.
    \label{eq_p0}
\end{equation}
Introducing another constant
\begin{equation}
    \eta = 4 \pi \frac{q^2}{m V_g} \xi > 0,
    \label{eq_eta}
\end{equation}
we transform eq.~\eqref{eq_p0} into the form
\begin{equation}
    \ddot{\vec{p}} = -\nabla_p \left(\eta \sqrt{1 + p^2}\right),
    \label{eq_p}
\end{equation}
where we used dot to denote time derivative. In what follows we use the following notations: superscript "+" denotes that the value is taken at the time instance $t + \Delta t$, the absence of such superscript indicates that the value is take at time instance $t$, and $\Delta$ is used to denote the difference, e.g. $\Delta \vec{p} = \vec{p}^+ - \vec{p} = \vec{p}\left(t + \Delta t\right) - \vec{p}(t)$ (in some cases we indicate the time instance explicitly to avoid confusion). 

As we can see \eqref{eq_p} is the equation that describes classical motion of an \textit{abstract} particle with coordinate $\vec{p}$ in a central potential $\eta \sqrt{1 + p^2}$. For the time interval $\Delta t$ this equation defines a definite solution $\vec{p}^+$, $\dot{\vec{p}}^+$ for arbitrary initial conditions:
\begin{equation}
    \vec{p} = \vec{p}(t), \ \dot{\vec{p}} = \frac{q}{m c} \sum_{j = 1}^M c_j \vec{E}_j.
    \label{p_ic}
\end{equation}

We consider the solution of \eqref{eq_p} in the next section, while here we complete the proof of consistency by showing that its solution unambiguously and consistently determine $\vec{E}_j^+$ and $\vec{r}^+$. To show this let us express the change of coordinate over the entire time step $\Delta \vec{r} = \vec{r}^+ - \vec{r}$ using eq.~\eqref{subsys_c4}:
\begin{equation}
    \Delta \vec{r} = \int_t^{t + \Delta t} \frac{c \vec{p}(\tau)} {\sqrt{1 + p^2(\tau)}} d \tau.
    \label{eq_Delta_r}
\end{equation}
Substituting \eqref{subsys_c3n} into \eqref{subsys_c2n} and integrating over the time step we get the change of $\vec{E}_j$:
\begin{equation}
    \Delta \vec{E}_j = \vec{E}_j^+ - \vec{E}_j = -4\pi c_j \frac{q}{V_g} \int_t^{t + \Delta t} \frac{c \vec{p}(\tau)} {\sqrt{1 + p^2(\tau)}} d \tau = c_j \Delta \vec{E},
    \label{eq_Delta_Ej}
\end{equation}
where we denote:
\begin{equation}
    \Delta \vec{E} = -\frac{4\pi q}{V_g} \Delta \vec{r}.
    \label{eq_Delta_E}
\end{equation}
Considering \eqref{subsys_c1n} at time instance $t + \Delta t$ we get:
\begin{equation}
    \dot{\vec{p}}^+ = \frac{q}{mc} \sum_{j=1}^M c_j \left( \vec{E}_j + c_j \Delta \vec{E}\right).
    \label{eq_dpdt_}
\end{equation}
Using \eqref{eq_Delta_r} -- \eqref{eq_dpdt_} we can express the change of field and the change of coordinate via $\dot{\vec{p}}^+$:
\begin{align}
    &\Delta \vec{E} = \xi^{-1}\left(\frac{mc}{q} \dot{\vec{p}}^+ - \sum_{j=1}^M c_j \vec{E}_j\right),\label{eq_dE}\\
    &\Delta \vec{E}_j = c_j \Delta \vec{E},\label{eq_dEj}\\
    &\Delta \vec{r} = -\frac{V_g}{4\pi q}\Delta \vec{E}.\label{eq_dr}
\end{align}
Thus our task is to find $\vec{p}^+$ and $\dot{\vec{p}}^+$ according to eq.~\eqref{eq_p}, preserving energy given by \eqref{eq_en_cons}. The solution of this task in combination with updates \eqref{eq_dE} -- \eqref{eq_dr} and the momentum update according to \eqref{subsys_b} can be referred to as \textit{energy-conserving particle-field pusher}. We consider one possible implementation of such pusher in the next section.

\begin{figure} 
\centering\includegraphics[width=1.0\columnwidth, scale=1]{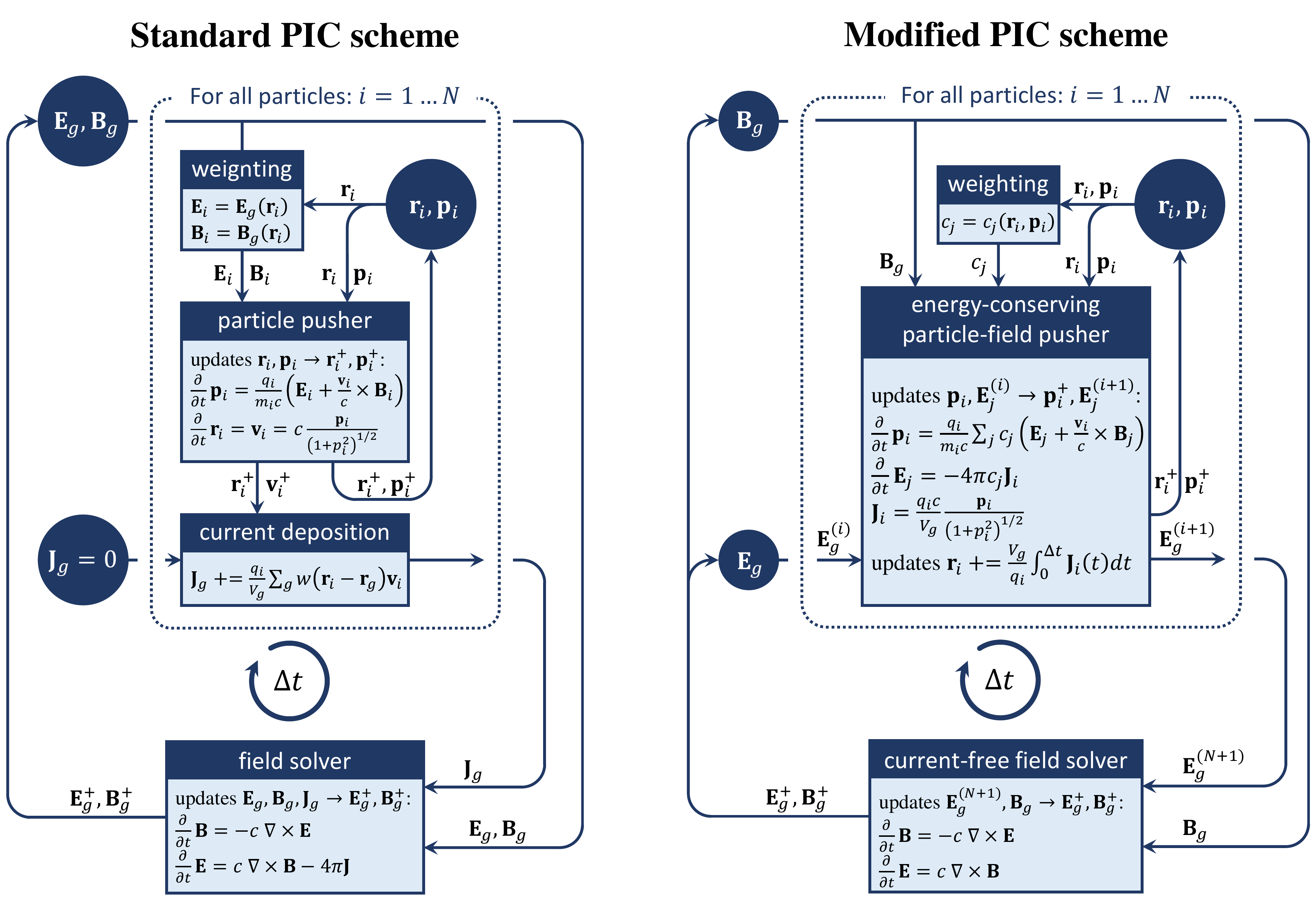}
\caption{The standard (left) and modified (right) PIC schemes. {For the modified scheme $\vec{E}^{(i)}$ denotes the electric field updated by the particle-field pusher applied for particles with indices up to $i - 1$.}}
\label{scheme}
\end{figure}

We conclude this section by summarizing the proposed modification of the PIC method. The method implies that we sequentially advance the state of each particle and of electric field at the coupled nearby nodes according to \eqref{subsys_c1n} -- \eqref{subsys_c3n}. Apart from this for each particle we advance the momentum according to the local magnetic field using rotation about the magnetic field vector. Finally, after advancing the states of all particles together with current-related change of the electric field at the coupled nodes, we use spectral solver to advance the field state according to Maxwell's equations without current. In fig.~\ref{scheme} we show the standard PIC scheme and the described modification in a way that highlights the differences: the circles are used to denote data, while the data processing principles are summarized in rectangular elements; the superscript "+" indicates the updated data; subscript "$g$" denotes grid values, $\vec{E}_g^{(i)}$ and $\vec{E}_g^{(i+1)}$ denote the state of electric field before and after processing the $i$-th particle, respectively.

\section{Explicit energy-conserving particle-field pusher}
\label{sec_coupling}

The energy-conserving particle-field pushed introduced in fig.~\ref{scheme} can be generally defined as an algorithm that advances the state of a given particle together with the state of the field at the nearby nodes according to both \eqref{subsys_b} and \eqref{subsys_c1} -- \eqref{subsys_c4}. An exact advancement warrants energy conservation given by eq.~\eqref{eq_en_cons}. For approximate solution we need to enforce energy conservation. Note that improved properties that reduce violation of energy conservation can be also practical because they can broaden the parameter region of numerical stability. However we here develop a solution with exact energy {conservation}.

Although a more comprehensive treatment can be useful, we here use several simplifications to get an approximate, yet energy-conserving solution of this problem. First, as mentioned above, we separate and split the effect of magnetic field, which we account for through the rotation of $\vec{p}$ about $\vec{B}$ before the advancement according to eqs.~\eqref{subsys_c1} -- \eqref{subsys_c4}; the details are given in the summary of the algorithm below. Next, we assume that coefficients $c_j$ are not changing in time during each time step. Under these assumptions, given relations \eqref{eq_dE} -- \eqref{eq_dr}, the remaining task is to advance the state of $\vec{p}(t)$ and $\dot{\vec{p}}(t)$ defined by \eqref{p_ic} over one time step according to \eqref{eq_p}:
\begin{equation}
    \ddot{\vec{p}} = -\nabla_p \left(\eta \sqrt{1 + p^2}\right), \ \eta > 0.
    \label{eq_p_particle_exact}
\end{equation}
This equation can be seen as an equation that describes motion of an \textit{abstract} non-relativistic particle with coordinate $p$ in a central potential $\eta \sqrt{1 + p^2}$. Hereafter we refer to this particle as \textit{p-particle}, so as not to confuse it with the real particle in question. 

The relation to the p-particle dynamics provides an insight into the origin of numerical heating and corresponding instability. In case of exact dynamics, p-particle is constantly being deflected towards the center, while in the standard PIC scheme the p-particle is advanced in time along a straight line pointing towards $\dot{\vec{p}}(t)$ (tangent to the trajectory at the initial point of the time step). In case of a small enough time step, the deviation is minor but always leads to p-particle being further from the center than it should be, meaning that the energy is constantly increased. We thus can conclude that accounting for the potential that restricts p-particle motion is the way to get rid of numerical {heating}.

For the motion of p-particle we have two invariants:
\begin{align}
    &\vec{p} \times \dot{\vec{p}} = L = const, \label{eq_pp_L}\\
    &\frac{1}{2}\left|\dot{\vec{p}}\right|^2 + \eta \sqrt{1 + p^2} = \mathcal{E} = const, \label{eq_pp_E}
\end{align}
where the first one represents the angular momentum conservation and the second one represent the energy conservation. Note that, despite having similar terms proportional to $\sqrt{1 + p^2}$, the energy conservation for p-particle differs from the energy conservation for the field and the real particle given by eq.~\eqref{eq_en_cons}. 

In case of eq.~\eqref{eq_p_particle_exact}, the standard way of determining particle dynamics in a central potential by adding centrifugal potential leads to differential equation \cite{jose.1998}:
\begin{equation}
    \dot{p} = \sqrt{2\mathcal{E} - L/p^2 - 2\eta\sqrt{1 + p^2}}.
\end{equation}
Since the exact solution is intricate, we choose to use another simplification with the following plan. We use another potential that approximates the one in eq.~\eqref{eq_p_particle_exact}, but leads to an explicit solution. We then have two main options to enforce exact energy conservation according to eq.~\eqref{eq_en_cons}. We can either use approximate $\vec{p}^+$ and correct the approximate $\Delta \vec{E}\left(\dot{\vec{p}}^+\right)$ (see eq.~\eqref{eq_dE}), or use approximate $\dot{\vec{p}}^+$, determine $\Delta \vec{E}\left(\dot{\vec{p}}^+\right)$ and then correct the approximate $\vec{p}^+$. Since the correction concerns 3D vector based on a single equation, there are many options. We choose to correct $\vec{p}^+$ keeping its direction unchanged, i.e. by multiplying vector $\vec{p}^+$ by a factor $\sigma$, which presumably should be close to unit. Since the correction of particle energy concerns the length of $\vec{p}$ only, the vector of difference between the approximate $\vec{p}^+$ and the corrected $\sigma \vec{p}^+$ has the minimal length among all possible ways of correcting $\vec{p}^+$ (determining such a minimal correction for $\Delta \vec{E}$ is also possible but requires some derivations). The condition for enforced energy conservation then takes the form: 
\begin{equation}
    \frac{V_g}{8\pi}\sum_{j=1}^M E_j^2 + m c^2 \sqrt{1 + p^2} = \frac{V_g}{8\pi}\sum_{j=1}^M {E_j^+}^2 + m c^2 \sqrt{1 + \sigma^2{p^+}^2}.
\end{equation}
We can explicitly express the correction factor $\sigma$:
\begin{equation}
    \sigma = \left(\left(\left(\frac{V_g}{8\pi m c^2} \sum_{j = 1}^M \left(E_j^2 - {E_j^+}^2\right) + \sqrt{1 + p^2}\right)^2 - 1\right){p^+}^{-2}\right)^{1/2},
    \label{eq_sigma}
\end{equation}

There is, however, an important restriction on the choice of approximate potential. It should restrict the motion of p-particle in such a way that the correction is always possible. In our case, it means that the resultant approximate $\dot{\vec{p}}^+$ doesn't lead to the field taking more energy than the particle can provide, otherwise we get imaginary value for $\sigma$. \\

\noindent \textbf{Lemma.} \textit{If $\dot{p}^+$ is limited by the maximal kinetic energy gain for p-particle then $\sigma$ is real.}\\

\noindent \textbf{Proof.} The restriction on the maximal kinetic energy gain for p-particle can be expressed as
\begin{equation}
    \left(\dot{p}^+\right)^2/2 \leq \eta \left(\sqrt{1 + p^2} - 1\right) + \dot{p}^2/2,
\end{equation}
which is achieved for $L = 0, \ p^+ = 0$. Let us now determine the upper bound for the electric field energy gain:
\begin{equation}
    \Delta \mathcal{E}_E = \frac{V_g}{8\pi} \sum_{j = 1}^M \left({E_j^+}^2 - E_j^2\right) = \frac{V_g}{8\pi}\sum_{j = 1}^M \left( 2 c_j \vec{E}_j \Delta \vec{E} + c_j^2 \Delta E^2\right),
    \label{eq_proof1}
\end{equation}
where we substituted $\vec{E}_j^+ = \vec{E}_j + c_j \Delta \vec{E}$ given eqs.~\eqref{eq_dE} -- \eqref{eq_dEj}. Since $c_j > 0$, the largest value is achieved in case all $\vec{E}_j$ are pointing along $\Delta \vec{E}$, therefore
\begin{equation}
    \Delta \mathcal{E}_E \leq \frac{V_g}{8\pi}\sum_{j = 1}^M \left( 2 c_j E_j \Delta E + c_j^2 \Delta E^2\right) = \frac{V_g}{8\pi} \left(2\Delta E \sum_{j = 1}^M c_j E_j + \xi \Delta E^2\right).
\end{equation}
Substituting eq.~\eqref{eq_dE}, we get
\begin{equation}
    \Delta \mathcal{E}_E \leq \frac{V_g m^2 c^2}{8\pi \xi q^2} \left(\dot{p}^+\right)^2 - \frac{V_g}{8\pi \xi} \left(\sum_{j = 1}^M c_j E_j\right)^2.
\end{equation}
Substituting restriction \eqref{eq_proof1} and expressing $\dot{p}^2$ via eq.~\eqref{subsys_c1n}, we obtain
\begin{equation}
    \Delta \mathcal{E}_E \leq m c^2 \left( \sqrt{1 + p^2} - 1\right).
    \label{ineq_proof1}
\end{equation}
Recalling the definition of $\Delta \mathcal{E}_E$ \eqref{eq_proof1} and substituting inequality \eqref{ineq_proof1} into eq.~\eqref{eq_sigma}, we can see that radicand is greater or equal to zero and thus $\sigma$ is real. \qedsymbol \\

Given the obtained restriction we choose to approximate the exact potential $\eta\sqrt{1 + p^2}$ by a parabolic potential $\kappa p^2/2$, where coefficient $\kappa$ is chosen to match the slope at the initial point $\vec{p}(t)$:
\begin{equation}
    \kappa = \frac{\eta}{\sqrt{1 + p(t)^2}}.
    \label{eq_kappa}
\end{equation}
This potential satisfies the identified requirement because the maximal drop of potential energy (reached at $p = 0$) is less than that of the exact potential. This follows from the fact that the gradient on the way to $p = 0$ is always smaller for the approximate potential:
\begin{equation}
    \forall p \leq p(t): \ \left|\nabla_p\left(\eta\sqrt{1 + p^2}\right)\right| \leq \left|\nabla_p\left(\kappa p^2/2\right)\right|.
\end{equation}
Note that the potentials exactly match in the limit $p^2 \ll 1$.

The equation for the dynamics of p-particle with the approximate potential takes the form of three dimensional symmetric harmonic oscillator:
\begin{equation}
    \ddot{\vec{p}} = - \kappa \vec{p}.
    \label{eq_p_particle_approx}
\end{equation}
To express the solution in explicit form we define a function
\begin{equation}
    \vec{g} = \dot{\vec{p}} + i \sqrt{\kappa}\vec{p}
\end{equation}
and compute its time derivative using \eqref{eq_p_particle_approx}:
\begin{equation}
    \dot{\vec{g}} = - \kappa \vec{p} + i \sqrt{\kappa} \dot{\vec{p}} = i \sqrt{\kappa} \vec{g}.
\end{equation}
Using the solution $\vec{g}^+ = \vec{g} \exp\left(i\sqrt{\kappa} \Delta t\right)$, we get
\begin{align}
    & \dot{\vec{p}}^+ = \operatorname{Re}\left\{\left(\dot{\vec{p}} + i \sqrt{\kappa}\vec{p}\right)\exp\left(i\sqrt{\kappa} \Delta t\right)\right\}, \label{eq_dotp_updated}\\
    & \vec{p}^+ = \frac{1}{\sqrt{\kappa}}\operatorname{Im}\left\{\left(\dot{\vec{p}} + i \sqrt{\kappa}\vec{p}\right)\exp\left(i\sqrt{\kappa} \Delta t\right)\right\}.\label{eq_p_updated}
\end{align}
Note that in case of non-relativistic description the approximate parabolic potential coincides with the exact one. In this case we do not need to do any correction and have exact energy conservation \eqref{eq_en_cons} already for \eqref{eq_dotp_updated} -- \eqref{eq_p_updated} with $\vec{E}_j^+$ determined according to \eqref{eq_dE} -- \eqref{eq_dEj}. In case of relativistic simulations, the solution \eqref{eq_dotp_updated} -- \eqref{eq_p_updated} for approximate potential becomes increasingly close to that for exact $\eta \sqrt{1 + p^2}$.

To complete the description of this basic form of energy-conserving particle-field pusher we address the question of changing the momentum according to the local magnetic field $\vec{B}$:
\begin{equation}
    \dot{\vec{p}} = \frac{q}{mc}\frac{\vec{p}}{\sqrt{1 + p^2}} \times \vec{B}.
    \label{eq_B_rotation}
\end{equation}
For constant magnetic field this equation describes the rotation of vector $\vec{p}$ about vector $\vec{B}$. Under this assumption we can express the solution using a rotation operator $\hat{R}_{\vec{a}}\vec{b}$, which rotates vector $\vec{b}$ about $\vec{a}$ by an angle $a$. To obtain explicit form of this operator we decompose $\vec{b}$ into the sum of parallel and perpendicular components with respect to vector $\vec{a}$:
\begin{align}
  &\vec{b} = \vec{b}_{||} + \vec{b}_{\bot},\\
  &\vec{b}_{||} = \vec{a} \frac{ \vec{b} \cdot \vec{a}}{a^2}, \
	\vec{b}_{\bot} = \vec{b} - \vec{a} \frac{\vec{b} \cdot \vec{a}}{a^2}.
\end{align}
Using this decomposition, we can express the rotation operator:
\begin{align}
& \hat{R}_{\vec{a}} \vec{b} = \vec{b}_{||} + \vec{b}_{\bot}\cos a + \frac{\vec{a} \times \vec{b}}{a}\sin a \nonumber\\
& = \vec{b} \cos a + \vec{a} \frac{\vec{b} \cdot \vec{a}}{a^2} \left(1 - \cos a\right) + \frac{\vec{a} \times \vec{b}}{a}\sin a.\label{eq_rotation_operator}
\end{align}
The rotation of $\vec{p}$ described by eq.~\eqref{eq_B_rotation} can be be given by
\begin{align}
    &\vec{p}^+ = \hat{R}_{\zeta(p)\vec{B}}\vec{p}, \label{p_b_update}\\
    &\zeta(p) = \frac{q \Delta t}{m c \sqrt{1 + p^2}}.
\end{align}
The rotation of momentum $\vec{p}$ due to magnetic field doesn't change the energy of the particle. We can apply it before or after coupling with the electric field. One can also split the step into two half-step rotations applied before and after coupling with the electric field. {Nevertheless, we leave possible improvements for further considerations elsewhere and use the simplest version of update based on the Boris pusher \cite{boris.1972}:
\begin{align}
    &\vec{t} = \frac{q \Delta t}{2mc\sqrt{1 + p^2}} \vec{B},\\
    &\vec{u} = \vec{p} + \vec{p}\times \vec{t},\\
    &\vec{p}^+ = \vec{p} + \frac{2}{1 + t^2} \vec{u} \times \vec{t},
\end{align}
and denote this algorithm of update by $\vec{p}^+ = \hat{R}^{Boris}_{\Delta t\vec{B}}\vec{p}$.}

We can now collect all the steps of such basic energy-conserving particle-field pusher, which for $i$-th particle converts {$\vec{r} = \vec{r}_i$,} $\vec{p} = \vec{p}_i$ and $\vec{E} = \vec{E}^{(i)}$ into {$\vec{r}^+ = \vec{r}_i^+$,} $\vec{p}_i^+ = \vec{p}^+$ and $\vec{E}^{(i+1)} = \vec{E}^+$ (see fig.~\ref{scheme}):

\begin{enumerate}
    \item $\displaystyle{c_j := W_{CIC}\left(\vec{r} + \frac{\Delta t}{2} \frac{c\vec{p}}{\sqrt{1 + p^2}}- \vec{r}^{(j)}\right), \ j = 1, ..., M}$
    \item $\vec{B} := \sum_j^M c_j \vec{B}_j, \tilde{\vec{p}} := {\hat{R}^{Boris}_{\Delta t\vec{B}}}\vec{p}$
    \item $\displaystyle{\xi := \sum_{j=1}^M c_j^2, \ \ \kappa := \frac{4\pi q^2}{m V_g \sqrt{1 + p^2}}} \xi$
    \item $\displaystyle{\dot{\tilde{\vec{p}}} := \frac{q}{m c} \sum_{j = 1}^M c_j \vec{E}_j}$\\
    $\displaystyle{\dot{\vardbtilde{\vec{p}}} := \operatorname{Re}\left\{\left(\dot{\tilde{\vec{p}}} + i \sqrt{\kappa}\tilde{\vec{p}}\right)\exp\left(i\sqrt{\kappa} \Delta t\right)\right\}}$\\
    $\displaystyle{\vardbtilde{\vec{p}} := \frac{1}{\sqrt{\kappa}}\operatorname{Im}\left\{\left(\dot{\tilde{\vec{p}}} + i\sqrt{\kappa}\tilde{\vec{p}}\right)\exp\left(i\sqrt{\kappa} \Delta t\right)\right\}}$
    \item $\displaystyle{\Delta \vec{E} := \xi^{-1}\left(\frac{mc}{q} \dot{\vardbtilde{\vec{p}}} - \sum_{j=1}^M c_j \vec{E}_j\right)}$\\
    $\displaystyle{\vec{E}_j^+ := \vec{E}_j + c_j \Delta \vec{E}, \ j = 1, ..., M}$
    \item $\displaystyle{\sigma := \left(\left(\left(\frac{V_g}{8\pi m c^2} \sum_{j = 1}^M \left(E_j^2 - {E_j^+}^2\right) + \sqrt{1 + p^2}\right)^2 - 1\right)\vardbtilde{p}^{-2}\right)^{1/2}}$\\
    $\displaystyle{\vec{p}^+ := \sigma \vardbtilde{\vec{p}}}$
    \item $\displaystyle{\vec{r}^+ := \vec{r} -\frac{V_g}{4\pi q}\Delta \vec{E}}$
\end{enumerate}

We conclude this section by outlining some limitations and possible improvements of the described algorithm. The presented consideration indicates two characteristic time scales of cyclic behaviour. The first one is induced by magnetic field and describes particle gyration with frequency $\omega_B = qB/mc\gamma$ (see eq.~\eqref{eq_B_rotation}), where $\gamma = \sqrt{1 + p^2}$ is the particle gamma factor. The frequency of the second one $\sqrt{\kappa}$ (see eq.~\eqref{eq_p_particle_approx}) corresponds to plasma frequency for the \textit{partial} density. To see this and clarify the meaning of the partial density, we recall that in PIC method the time evolution is computed for macroparticles that are assumed to represent a number of real particles, the number of which is commonly referred to as \textit{weight}. Denoting the real particle mass and charge for the $i$-th macroparticle with weight $w_i$ by $q_r = q_i/w_i$ and $m_r = m_i/w_i$ respectively, we can represent the frequency $\sqrt{\kappa}$ in the following form (see eqs.~\eqref{eq_kappa} and \eqref{eq_eta}):
\begin{equation}
    \sqrt{\kappa} = \sqrt{\frac{4\pi\left(w_i q_r\right)^2}{w_i m_r \gamma V_g}\xi} = \sqrt{\frac{4\pi q_r^2 n^p_i}{m_r\gamma}},
    \label{eq_eff_plasma_freq}
\end{equation}
where partial density $n_i^p = w_i/(V_g/\xi)$ corresponds to the density that real particles associated the $i$-th macroparticle would have in case of even distribution in an effective volume $V_g/\xi$ (note that $\xi \sim 1$ for the CIC weighting). The right-hand side of eq.~\eqref{eq_eff_plasma_freq} indicates that $\sqrt{\kappa}$ represents the time scale of plasma oscillations with the density of $n_i^p$ and with account for the relativistic mass increase.

The algorithm described in this section provides the state advancement that can be sufficient for simulations even if the outlined time scales of cyclic behaviour locally become comparable to or even shorter than the time step. For example, the algorithm can be sufficient if the accurate tracking of the phases of oscillations is not important and the energy conservation is a sufficient property that provides an appropriate modelling of particle's response for the simulation of a macroscopic processes in question. Nevertheless, if a more accurate treatment is needed, there are few interesting possibilities for improvement. To reveal such possibilities let us consider in what ways the algorithm can be inaccurate.

First, a poor resolution of particle gyration in magnetic field may lead to inaccurate description of the particle drift and energy gain near stopping points during direct laser acceleration (see \cite{arefiev.pop.2015, tangtartharakul.jcp.2021}). To mitigate these effects in relativistic simulations one can apply the approach of \textit{individual subcycling} proposed by Arefiev et al. \cite{arefiev.pop.2015} and further developed in \cite{tangtartharakul.jcp.2021}. In the context of described algorithm the individual subcycling means that for each particle we determine an individual time sub-step $\Delta t_i$ so that $\omega_B \Delta t_i \ll 1$, and repeat steps 1-3 with this time sub-step (if $\Delta t_i \geq \Delta t$ we do a single time step with $\Delta t$). 

The individual subcycling can also provide a way to account for the changing gamma factor during plasma oscillation. This is related to another source of inaccuracy that is caused by the use of approximate parabolic potential instead of $\eta\sqrt{1 + p^2}$. To mitigate this inaccuracy the choice of individual sub-step should be driven by requiring $\Delta t_i \sqrt{\kappa} \ll 1$.

{The proposed use of parabolic approximation for the potential in eq.~\eqref{eq_p} becomes increasingly accurate for low energies (i.e for $p^2 \ll 1$). That is why the approximation becomes negligible for plasma states with non-relativistic temperatures, while still being reasonable for relativistic dynamics, particularly in the sense of restricting the gain of kinetic energies for particles in case of large time steps. Let us note that the Euler-type momentum advancement in a given electric field (used in the Boris method) can be seen as the replacement of $\eta\sqrt{1 + p^2}$ potential with zero potential, in which the p-particle propagates linearly according to the initial $\dot{\vec{p}}$ determined by eq.~\eqref{p_ic}. In this context the proposed parabolic approximation appears to be a good tradeoff between accuracy and computational complexity: instead of developing a scheme with higher order accuracy or even exact treatment of \eqref{eq_p} it might be more beneficial to use larger number of particles and/or perform collective subcycling.}

One other source of inaccuracy is related to the fact that the simulated plasma frequency \eqref{eq_eff_plasma_freq} corresponds to partial density of a single macroparticle, while a large number of macroparticles for the given space location would imply a higher plasma frequency. This kind of inaccuracy is related to the fact that particles interact via field, not directly. To mitigate this effect, one can apply the following procedure. For each particle we determine a time sub-step that ensures a good resolution of the actual plasma oscillations 
\begin{equation}
\Delta t_i \ll \left(\frac{4\pi q_r^2 n_r}{m_r \gamma}\right)^{-1/2},    
\end{equation}
where $n_r$ is the estimated local density of particles. If $\Delta t_i < \Delta t$ the particle is advanced in time by $\Delta t_i$ and flagged as a particle with uncompleted time step with individual time $t_i = t + \Delta t_i$. Once all particles are processed, we return back to the flagged particles and advance each again by another individual time sub-step, increasing $t_i$ by $\Delta t_i$ and unflagging it if $t_i = t + \Delta t$ (the substep $\Delta t_i$ can either be adjusted to fit a whole number of sub steps in $\Delta t$ or the last step can be cut to fit the time at the end). We repeat this cycle over flagged particles until they all become unflagged. To highlight the difference in the way of processing particles one can call this \textit{collective subcycling}. It is important that the procedure becomes computationally demanding only locally, where the density is high. Nevertheless, accounting for multiple plasma oscillations within a single time step may still be inaccurate because we do not account for the electromagnetic field evolution during a single time step. Effectively, we inaccurately describe the oscillating current that would otherwise result in energy emission in the form of electromagnetic waves. In a more general case, the plasma oscillations can be strongly coupled with electromagnetic waves. That is why there should be some practical limit for the use of subcycling.

One other reason for resolving the time scale of $\sqrt{\kappa}$ is related to the motion of particles. To see the mechanism of this limitation, let us consider the case, when the time step is so large that we have $\sqrt{\kappa}\Delta t = \pi$ and also let us assume that $\vec{p} = 0$. In this case, $\Delta\vec{E} = 2\vec{E}$ no matter how strong the field is (one can interpret this as a half oscillation for plasma associated with a single particle). Following step 7 of the algorythm this means that the change of coordinate $\vec{r}^+ - \vec{r} = V_g E/2\pi q$ can potentially exceed $c \Delta t$. One can interpret this limitation as the necessity to resolve plasma period for partial density created by individual particles. Assuming equal weights of particles, this period is larger than the actual plasma period by a factor that is equal to the square root of the number of particles per cell. 

There is a source of inaccuracy that is related to the discrepancy between the $c_i$ determined based on the "old" velocity and the $c_i$ that correspond to the actual change of particle coordinate $\Delta \vec{r}$. This is related to the violation of the continuity equation for the current and thus can lead to the accumulation of spurious charge (note that the correction based on solving Poisson's equation is possible but can violate the energy conservation). This kind of inaccuracy becomes increasingly large with more localized shape functions and when the motion is relativistic (and also when the particle is near the stopping points). Although subcycling can also alleviate this difficulty, there is another possibility. In case the determined $\Delta \vec{r}$ corresponds to largely different $c_i$, we can repeat the computations with $c_i$ determined at $\vec{r}+\Delta \vec{r}/2$. One can even run a cycle of such correcting procedure, expecting that it quickly converges to a consistent advancement.  

Finally, there is one other collective source of inaccuracy related to the fact that during each step the particles interact with the field sequentially. This means that field experienced by each particle depends on the order of processing. The effect is small if the time step is small, as is the total field change over each time step. Nevertheless, it is interesting to consider the possibility of using large time steps. In this case, one can again use local collective subcycling. However, there is also another simple measure to suppress the effect of order. Since the order of processing can be correlated with the particle location in phase space, ordered processing means that different parts of phase space are repeatedly experiencing the fields at the same stage of update after the update by the field solver. To remove the accumulation of such bias we can shuffle the order of processing. This is implemented in the provided implementation, while the positive effect of shuffling is shown in Sec.~\ref{sec_cost}.

\section{Field advancement using spectral method}
\label{sec_spectral_solvers}

To complete the time step so that the energy is exactly preserved we need to do energy-conserving advancement of electromagnetic field. This can be done by standard spectral method that in addition is free of numerical dispersion. Although the spectral method can be considered a well-recognized alternative to finite-difference and finite-element methods (i.e. grid-based methods) it remains a rather uncommon choice in many numerical studies. We thus provide a brief description of the method for completeness.

The spectral solver of Maxwell's equations was described by Haber \textit{et al.} in 1973 \cite{haber.cnsp.1973}, by Buneman \textit{et al.} in 1980 \cite{buneman.jcp.1980}, and later by Birdsall and Langdon in their textbook "Plasma Physics via Computer Simulation" (see chapter 15.9 (b)) published in 1984 \cite{birdsall.1984}. The solver implies that Fast Fourier Transform (FFT) quickly converts the field defined on a grid into a sum of plane wave modes, each of those is independently advanced in time without errors, and then returns the field to the initial coordinate representation on the grid. This gives exact solution (without numerical dispersion) in case of constant current. In 1997 Liu considered an approximate/truncated version called pseudo-spectral time-dependent (PSTD) \cite{liu.motl.1997}, which doesn't require sine/cosine function computation but is subject to numerical dispersion. In 2013 Vay \textit{et al.} considered the exact solution (also for time staggered field components) and introduced the term pseudo-spectral analytical time-domain (PSATD), which highlights the difference from PSTD \cite{vay.jcp.2013}. In the same work Vay \textit{et al.} proposed a solution to the problem of arranging parallel computations being the main stumbling block for the broad use of the method. Exploiting linearity of Maxwell's equation and the finiteness of the signal propagation speed, the method in parallel advances the field in spatial domains having overlaps, where the cross-summation of the fields is carried out once in a time interval less than that needed for the light to traverse the overlap. A different method for arranging parallel computations, as well as the PIC code ELMIS based on this method, was described and used for numerical studies in PhD thesis \cite{gonoskov.phd.2013} also published in 2013. The method is based on alternating data decomposition throughout each computational step (for details see Sec.~\ref{sec_implementation}). Although the term PSATD is being used in recent works \cite{lehe.cpc.2016, fedeli.cs.2022}, we here revert to the original term \textit{spectral solver} used in the pioneering works \cite{haber.cnsp.1973, buneman.jcp.1980, birdsall.1984}.

The spectral solver for Maxwell's equations can be seen as a particular case of a general approach to numerical solution of partial differential equations (PDE). The idea is to use time step splitting so that the time advancement due to linear part (with optional inclusion of constant terms) of a PDE is carried out using eigenbasis so that the time-advancing operational exponent is computed exactly via the Jordan canonical form. In physics the relevant bases for getting useful diagonal representation of the operational exponent correspond to wave and coordinate representations of the state function, which means that the method can be greatly facilitated by the use of FFT. Although general procedure is applicable to the case of Maxwell's equations, it is instructive to consider the derivation of the spectral methods in physical terms. Following Ref.~\cite{buneman.jcp.1980}, we consider complex-value field that combines electric (real part) and magnetic (imaginary part) field components:

\begin{equation}\label{F_def}
{\bf F} = {\bf E} + i{\bf B},
\end{equation}
where $i$ is the imaginary unit. For vector $\vec{F}$ Maxwell's equations with zero current take the following form:
\begin{align}
		& \nabla \times {\bf F} = \displaystyle{\frac{i}{c}\frac{\partial}{\partial t}{\bf F}},	\label{F_Maxwell1}\\
		& \nabla \cdot {\bf F} = 4 \pi \rho, \label{F_Maxwell2}
\end{align}
where $\rho$ is the charge density and the current $\vec{J}({\bf r})$ is omitted in accordance with the PIC modification in question. Applying Fourier transform to both sides of equations (\ref{F_Maxwell1}, \ref{F_Maxwell2}) and denoting the representations of functions ${\bf F}({\bf r})$ and ${\rho}({\bf r})$ in the basis of wave-vectors ${\bf k}$ by subscript ${\bf k}$, we get: 
\begin{align}
		& \displaystyle{\frac{1}{c}\frac{\partial}{\partial t}{\bf F_{\bf k}}} =  \displaystyle{{\bf k} \times {\bf F_{\bf k}}},\label{F_Maxwell_k1}\\
		& {\bf k} \cdot {\bf F_{\bf k}} = -4 \pi i \rho_{\bf k}.\label{F_Maxwell_k2}
\end{align}
Eq.~\eqref{F_Maxwell_k2} corresponds to Poisson’s equation and remains true if the current and charge fulfill the continuity equation. The propagation of electromagnetic radiation is described by eq.~\eqref{F_Maxwell_k1}, which describes the rotation of vector $\textbf{F}_{\bf k}$ about vector $\textbf{k}$ with frequency $c k$. Note that one can also exactly solve the equation for $\vec{F}$ with constant current (see \cite{birdsall.1984}), but this is not of interest for us because the current is already accounted for in the particle-field pusher. We thus can express the time advancement via rotation operator given by eq.~\eqref{eq_rotation_operator}: 
\begin{equation}
    \vec{F}^+ = \hat{R}_{c\Delta t \vec{k}} \vec{F}.
\end{equation}

\section{{Accuracy order and possible improvements}}
\label{sec_acc_improvements}

{In this section we demonstrate that the described implementation has only the first order accuracy in time, elaborate a variant of the method that has the second order accuracy and consider other possibilities for improvements. To assess the accuracy order we consider separately three distinctly different aspects.}

{The first one concerns coupling between the electric field and the momentum of each particle. In the conventional Boris pusher this coupling is not included in the momentum update in the sense that the field is assumed to remain unchanged during the updates (the coupling is enabled via current). This becomes an increasingly accurate assumption with decrease of time step and particle weight (increase of the number of particles per cell). The time staggered update of momentum and field results in the second order accuracy. However, the proposed energy-conserving method aims at handling large time steps and thus we here need to consider a more complicated problem that includes the coupling with the field.}

{The second aspect concerns the use of parabolic approximation of the potential in eq.~\eqref{eq_p}, which can be also interpreted as the use of an effective relativistic mass increase. For the particle-field coupling the first order accuracy for the update of momentum $\vec{p}$ corresponds to solving equation $\ddot{\vec{p}} = 0$, which takes into account only instantaneous electric field state at the time instance of update. Matching the gradient of the potential in eq.~\eqref{eq_p} with the parabolic potential leads to the first order accuracy for the update of $\dot{\vec{p}}$ and the second order accuracy for the update of  $\vec{p}$.}

{The third aspect concerns splitting of the ensemble update into a sequence of updates for all the particles separately. This can be seen as a multiple application of the Strang splitting \cite{strang.jna.1968, mclachlan.an.2002} to the coupled system of field and all the particles. To assess the accuracy order and the possibility for its increase let us consider the case of just two particles interacting with the field. Since we already considered the aspect of approximating relativistic potential with an effective non-relativistic potential, we assume here that both particles are non-relativistic. For simplicity we also assume that the particles have the same charges and masses, and are coupled with a single field node. Eqs.~\eqref{subsys_c1} -- \eqref{subsys_c3} for such a simplified case take the form:
\begin{align}
    & \frac{\partial\vec{p}_1}{\partial t} = \alpha \vec{E},\\
    & \frac{\partial\vec{p}_2}{\partial t} = \alpha \vec{E},\\
    &\frac{\partial\vec{E}}{\partial t} = \beta \left(\vec{p}_1 + \vec{p}_2\right),\\
\end{align}
where $\alpha = q/mc$, $\beta = -4\pi q c$. This system can be written in the matrix form:
\begin{equation}
    \frac{\partial}{\partial t} \psi = S \psi,
\end{equation}
where the state vector $\psi$ and matrix $S$ are defined by:
\begin{equation}
    \psi = \begin{bmatrix}\vec{p}_1\\ \vec{p}_2\\ \vec{E}\end{bmatrix}, \:\:
    S = \begin{bmatrix}0 & 0 & \alpha \\0 & 0 & \alpha \\ \beta & \beta & 0 \end{bmatrix}.
\end{equation}
The exact solution to this problem for updating the state from $\psi(t)$ to $\psi(t + \tau)$ has the form:
\begin{equation}
    \psi(t + \tau) = e^{\tau S} \psi(t).
    \label{S_ex}
\end{equation}
The idea of Strang splitting is based on representing matrix $S$ as a sum of two matrices $S_1$ and $S_2$, each of those describe a subsystem with reduced complexity. In our case the matrices take the form:
\begin{equation}
    S = S_1 + S_2, \:\:
    S_1 = \begin{bmatrix}0 & 0 & \alpha \\0 & 0 & 0 \\ \beta & 0 & 0 \end{bmatrix}, \:\:
    S_2 = \begin{bmatrix}0 & 0 & 0 \\0 & 0 & \alpha \\ 0 & \beta & 0 \end{bmatrix}.
\end{equation}
As one can see the subsystem described by $S_1$ concerns the update of $\vec{p}_1$ and $\vec{E}$, whereas the subsystem described by $S_2$ concerns the update of $\vec{p}_2$ and $\vec{E}$. Solving these systems sequentially for a small time step $\tau$ can be seen as the main idea behind the proposed method labeled below with subscript "ec":
\begin{equation}
    \psi(t + \tau) \approx \psi_{ec}(t + \tau) = e^{\tau S_1} e^{\tau S_2} \psi(t).
    \label{S_ec}
\end{equation}
To determine the accuracy order related to splitting we consider the difference between the exact solution $\psi(t + \tau)$ and the approximate solution given by $\psi_{ec}(t + \tau)$:
\begin{equation}
    \psi(t + \tau) - \psi_{ec}(t + \tau) = \left(e^{\tau\left(S_1 + S_2\right)} - e^{\tau S_1} e^{\tau S_2}\right)\psi(t) = \frac{\tau^2}{2}\left(S_2 S_1 - S_1 S_2\right) + O\left(\tau^3\right).
\end{equation}
Since matrices $S_1$ and $S_2$ do not commute, the solver based on the introduced splitting has the first order accuracy.}

{The presented analysis indicate a possibility for increasing the accuracy order. Let us first split the exact solution \eqref{S_ex} into two equal sub-steps and then apply splitting for both sub-steps, but using the reversed order in eq.~\ref{S_ec} for the second sub-step:
\begin{align}
    e^{\tau S}& = e^{\frac{\tau}{2}S} e^{\frac{\tau}{2}S} = e^{\frac{\tau}{2}S_1 + \frac{\tau}{2}S_2} e^{\frac{\tau}{2}S_2 + \frac{\tau}{2}S_1} \\
    & = \left( e^{\frac{\tau}{2}S_1} e^{\frac{\tau}{2}S_2} + \frac{\tau^2}{8}\left(S_2 S_1 - S_1 S_2\right) + O\left(\tau^3\right)\right)
    \left( e^{\frac{\tau}{2}S_2} e^{\frac{\tau}{2}S_1} + \frac{\tau^2}{8}\left(S_1 S_2 - S_2 S_1\right) + O\left(\tau^3\right)\right) \\
    & = e^{\frac{\tau}{2}S_1} e^{\frac{\tau}{2}S_2} e^{\frac{\tau}{2}S_2} e^{\frac{\tau}{2}S_1} + O\left(\tau^3\right).
\end{align}
Thus, the following expression describes the solver with the second order accuracy, which is labeled with "ec2" below: 
\begin{equation}
    \psi_{ec2}(t + \tau) = e^{\frac{\tau}{2}S_1} e^{\tau S_2} e^{\frac{\tau}{2}S_1} \psi(t).
    \label{S_ec2}
\end{equation}
Eq.~\eqref{S_ec2} indicates a way to handle multiple subsystems. For instance, if we have three subsystems $S = S_1 + S_2 + S_3$ we can first apply this expression with respect to two subsystems $S_1$ and $S_2 + S_3$ and get the solution with the second order accuracy expressed via $e^{\tau (S_2 + S_3)}$. We then can apply this expression to split the system $S_2 + S_3$, which leads to the following solver with the second order accuracy:
\begin{equation}
    \psi_{ec2}(t + \tau) = e^{\frac{\tau}{2}S_1} e^{\frac{\tau}{2}S_2} e^{\tau S_3} e^{\frac{\tau}{2}S_2} e^{\frac{\tau}{2}S_1} \psi(t).
\end{equation}
Iterative application of this splitting leads to the method modification that has the second order accuracy for $N$ particles:
\begin{align}
    & S = \sum_i S_i,\\
    & \psi_{ec2}(t + \tau) = e^{\frac{\tau}{2}S_1} e^{\frac{\tau}{2}S_2} ... e^{\frac{\tau}{2}S_N} e^{\frac{\tau}{2}S_N} ... e^{\frac{\tau}{2}S_2} e^{\frac{\tau}{2}S_1}\psi(t).
\end{align}
In practice this means that we first do a loop for updating all the particles in any random order  by a half-step and then do another half-step update making a loop in the exactly reversed order. Note that the second order solver ("ec2") requires roughly twice as much time for update as compared to the first order solver ("ec"). Nevertheless, as we demonstrate in Sec.~\ref{sec_limit} the increased order of accuracy can make the results much more accurate than that obtained with twice as many steps and the first order solver.}

{Let us also note that the presented consideration concerns only the coupling itself, while the use of estimated mid-point of the trajectory element for determining coupling coefficients introduces another error related also to the space step. For example, the described estimate becomes increasingly inaccurate if particles undergo cyclotron gyration with a poorly resolved period. The effect of this error depends on the simulated process: the case of low particle energy and strong magnetic fields can be problematic. As was mentioned, individual subcycling and correction of the initial estimated mid-point location can provide an improvement in this case.}

{We now turn to the consideration of two other potential improvements. The first one concerns the violation of charge conservation known to be a severe problem in some simulations. One well-known solution to this problem is Esirkepov's charge conserving current deposition \cite{esirkepov.cpc.2001}, which computes the current produced by the change of location for each particle in an exact agreement with the charge continuity equation. Since in our case the current is not computed, we cannot apply this method. At the same time, in our implementation there are no measures that could ensure an exact agreement with the charge continuity equation. For sufficiently large steps the errors can be accumulated, which can lead to numerical artefacts. We demonstrate the existence of this drawback in the test presented in Sec.~\ref{sec_verification}. However, since we use the Fourier-based method for updating electromagnetic fields we can apply the procedure of divergence cleaning. For this we exploit the fact that exact charge conservation corresponds to maintaining the exact solution of Poisson's equation described by eq.~\eqref{F_Maxwell_k2}. We can thus compute charge density $\rho$ making a loop over all particles, transform it into the bases of plane waves $\rho_\vec{k}$ using additional FFT and then correct vector $\vec{F}_\vec{k}$ in accordance with eq.~\eqref{F_Maxwell_k2}. A straightforward way of such correction is to replace the projection of vector $\vec{F}_\vec{k}$ on vector $\vec{k}$ with the exact value:
\begin{equation}
    \vec{F}_\vec{k} := \vec{F}_\vec{k} - \frac{\vec{k}\cdot\vec{F_\vec{k}}}{k} \frac{\vec{k}}{k} - 4\pi i \rho_\vec{k} \frac{\vec{k}}{k}.
\end{equation}
Nevertheless, this procedure does not preserves energy and thus deprives our method of its characteristic property. The example presented in Sec.~\ref{sec_verification} shows that the proposed energy conservation scheme does not necessarily run into a complete conflict with divergence cleaning, but the disagreement can be essential. That is why this modification is left as an auxiliary option that does not necessarily lead to an improved accuracy of simulations.}

{There is however a possibility of exploiting the proposed splitting for combining the properties of both charge and energy conservation. The schematic description of this possibility is as follows. For each particle we restrict the change of electric field at the nearby nodes in accordance with charge continuity equation, quantifying the extent of change by one or several parameter (optionally using particle's instantaneous direction of motion). Then we introduce a sub-system that couples particle’s momentum and the field change quantified by these parameters. The solution of this system can provide a way to either exactly preserve energy and charge or make a significant improvement to the method described in this paper.}

{The second potential improvement concerns the possibility of preserving momentum. Instead of using the part of Boris pusher related to magnetic field (or the  particle momentum rotation described by eq.~\eqref{p_b_update}) for solving eq.~\eqref{subsys_b} we can express the corresponding part of momentum update $\Delta_\vec{B} \vec{p}$ via the change of electric field $\Delta \vec{E}$ using eqs.~\eqref{eq_Delta_r}, \eqref{eq_Delta_E}:
\begin{align}
    \Delta_\vec{B} \vec{p} & = \int_t^{t + \Delta t} \frac{q}{mc} \frac{\vec{p}}{\sqrt{1 + p^2}}\times \vec{B} = \frac{q}{mc} \Delta \vec{r} \times \vec{B}
    = -\frac{V_g}{4\pi mc} \Delta \vec{E} \times \vec{B}.
\end{align}
Expressing the corresponding change of particles' momentum in dimensional units
\begin{equation}
    mc \Delta_\vec{B} \vec{p} = - V_g \frac{c}{4\pi} \left(\sum_{j = 1}^M c_j \Delta \vec{E} \right) \times \vec{B} = \sum_{j = 1}^M V_g \frac{c}{4\pi} \Delta \vec{E}_j \times \vec{B},
\end{equation}
we see that this change of momentum is exactly equal to the change of momentum carried by the electromagnetic field. This can provide the property of momentum conservation. However, in contrast to the use of Boris pusher or eq.~\eqref{p_b_update} this update does not necessarily preserves particle's energy and thus its use would deprive our method of its characteristic property. Nevertheless, this observation indicates an interesting pathway towards further developments that exploit the described splitting and particle-field coupling. One possible direction is to consider not exact but improved properties with respect to preservation of charge, energy and momentum. The other option is to consider a family of solvers that can prioritize the conservation of one or another quantity.}

\section{\texorpdfstring{$\pi$}{pi}-PIC}
\label{sec_implementation}
Here we briefly describe an open-source {python library} $\pi$-PIC (PI-PIC: Python-controlled Interactive PIC), {which provides implementations of the described PIC solvers \cite{pipic}.} The developments are done within the hi-$\chi$ framework, which is an open-source collection of tools for performing simulations and data analysis in strong-field particle and plasma physics \cite{hichi}. The code is designed for relativistic simulations in 3D, 2D and 1D geometry. The computational routine is based on spectral field solver and includes a basic implementation of the described energy conservation concept (mid-point CIC coupling) {with the first and second order accuracy,} along with the standard PIC method with Boris pusher. The inclusion of {several} solvers is to facilitate comparison and studies of the proposed approach. Simulations can be configured and executed using Python (or Jupyter notebook). 

The code is written in C++ and resource intensive parts exploit OpenMP. For each type particles are stored in separate vectors (operated by standard library \texttt{vector}) allocated for each cell. To make possible use of multiple threads the cells are processed with {a stride of 8 along $x$ axis}, which in case of CIC coupling/weighting ensures thread-safe operation {(note that the estimated mid-point for a particle in a given cell can be in a neighboring cell). To implement shuffling, in the beginning of each iteration we chose a random order of offsets along $x$ axis (from 0 to 7) and make a non-parallel loop over all the offset values. For each offset we use OpenMP to make a parallel loop over groups of cells: each group includes all the cells with the same location along $x$ axis; the groups are shifted by the given offset and spaced apart by the value of stride along $x$ direction. The loop over all cells along $y$ and $z$ (in case of 2D or 3D geometry) are made by the same thread.} If a particle migrates to a neighboring cell it is thread-safe to remove it from the current vector and add to the corresponding vector of the neighboring cell (migrated processed particles are stored in the demarcated end part of the vector to avoid double processing). If a particle migrates to a {distant} cell, the reference to this particle is added to thread-local list of particles to be moved after the parallel loop over particles (note that in this case CIC might be insufficient and one should consider reducing time step or using more advanced coupling method). {In addition, in case of energy-conserving solver the particles are shuffled in each cell to remove repeated effect of ordered interaction with the field (for more details see Sec.~\ref{sec_cost}).} The implementation of the energy-conserving solver with the second order accuracy is based on making two identical loops with the exactly reversed order (to maintain the exactly reversed order the particles are note migrated during the first loop). This was measured to increase the computational demands by {about} a factor of two.

{$\pi$-PIC is developed as a Python library, which offers a possibility to configure data containers for fields and particles. All data elements can be accessed for output and advanced in time using} one of the following {solvers (one can also assign external callbacks (extensions) developed in Python, C++ or other languages to be applied during particle loops)}:
\begin{itemize}
    \item "\texttt{ec}" stands for the described above simplest version of energy-conserving solver with the first order accuracy and coupling coefficients determined by the CIC weighting for the predicted mid-point location;
    \item "\texttt{fourier\_boris}" stands for the standard Boris pusher with CIC weighting combined with the Fourier-based solver of Maxwell's equations, divergence cleaning and time-centered application of current, i.e. half-step effect before and after field update;
    \item "\texttt{ec2}" stands for the variant of energy-conserving solver that has the second order accuracy reached by making two loops over particles with mutually reversed order of processing (see sec.~\ref{sec_acc_improvements}).
\end{itemize}
In addition one can disable divergence cleaning for the Boris method {using the command\\
"\texttt{fourier\_solver\_settings(divergence\_cleaning = False)}".} In what follows this option is labeled "\texttt{boris\_no\_div\_cleaning}".

{Although the code optimization is beyond the scope of this paper, we here report some basic measurements and observations for future reference. We consider the simulation of laser-solid interaction (see Sec.~\ref{sec_verification} for description and Appendix.~\ref{source_verification} for the Python code), using 2D grid with 100 particles per cell in the plasma bulk and the rate of particle migration between cells ranging from 0.06 to 0.2 (0.5 in case of Boris pusher due to numerical heating). When using 8 cores of Intel Core i7-10700K@3.8GHz the average update time per particle ranges from 12 to 14 ns for \texttt{fourier\_boris} solver, while for \texttt{ec} and \texttt{ec2} it ranges from 10 to 11 and from 19 to 20 ns, respectively. The update time with \texttt{ec} appears to be even shorter than that with \texttt{fourier\_boris}, which can be due to additional memory calls for updating grid values of charge density and current in the latter case. Compared to these time costs, shuffling particles does not take a significant amount of time, probably due to good use of caching.}

Computations related to spectral field solver are also arranged using OpenMP loop over individual 1D FFT calls. The time demands should be attributed to the used FFTW library \cite{frigo.ieee.2005}. Nevertheless we note that the described method doesn't require FFT for the current. For reference, for the above mentioned CPU the best time for $64\times64\times64$ was measured to be {10} ns per cell.

{We should emphasize that the described strategy for parallel processing is significantly restricted as compared to the possibilities for the conventional PIC method and might be inefficient for a large number of cores and especially for GPU. The parallel implementation for distributed memory of supercomputers can be done using decomposition into domains processed by each supercomputer node. However, in contrast to conventional PIC computations the proposed algorithm would require transferring electric field values at the bordering nodes for alternating processing of cells at the interfaces. That is why the order of offsets must be synchronized. The implementation of the algorithm with the second order accuracy would require making an additional loop with the inversed order of processing.}

The controls of $\pi$-PIC are rather minimalistic, yet they can be found sufficient for many cases. To maintain reasonable performance the communication between the data of simulations and Python is arranged using C callbacks provided by \texttt{Numba} package. This approach has been previously developed and used in hi-$\chi$ framework \cite{panova.as.2021}. From Python it is possible to call a full cycle over all cells or particles so that for each particle or cell a function defined in Python is called. Along with the address of that function, it is possible to pass pointers to arbitrary arrays of integer or double type to be used or amended during the function calls. The use is exemplified in the next subsection.

The interface is based on several assumptions:
\begin{itemize}
    \item CGS units are used for all dimensional variables;
    \item the components of vectors are enumerated: $x$, $y$, $z$ {correspond} to [0], [1], [2], respectively;
    \item the sizes of the grid \texttt{nx}, \texttt{ny} and \texttt{nz} should be powers of two to facilitate FFT; 
    \item 2D simulation is enabled by setting \texttt{nz=1}, 1D simulations is enabled by \texttt{nz=1} and \texttt{ny=1};
    \item the space is assumed to be periodic (other geometries can be implemented within Python).
\end{itemize}

\subsection{{Example: plasma oscillation}}
\label{sec_example}

Here we provide a Python file that illustrates the use of $\pi$-PIC {(see \cite{pipic} for more examples and configuration files used for simulations in the manuscript)}. The case in question is the process of plasma oscillation.

\begin{minted}[fontsize=\small,xleftmargin=20pt,linenos]{python}
import pipic
from pipic import consts, types
import matplotlib.pyplot as plt
import numpy as np
from numba import cfunc, carray
import os, time
from pipic.consts import electron_mass, electron_charge, light_velocity

# ===========================SIMULATION INITIALIZATION===========================
temperature = 1e-6*electron_mass*light_velocity**2
density = 1e+18
debye_length = np.sqrt(temperature/(4*np.pi*density*electron_charge**2))
plasma_period = np.sqrt(np.pi*electron_mass/(density*electron_charge**2))
l = 128*debye_length
xmin, xmax = -l/2, l/2
field_amplitude = 0.01*4*np.pi*(xmax-xmin)*(-electron_charge)*density
nx = 128
time_step = plasma_period/64

# ---------------------setting solver and simulation region----------------------
sim = pipic.init(solver='ec', nx=nx, xmin=xmin, xmax=xmax)

# ------------------------------adding electrons---------------------------------
@cfunc(types.add_particles_callback)
def density_callback(r, data_double, data_int):
    return density

sim.add_particles(name='electron', number=1000*nx,
                  charge=electron_charge, mass=electron_mass,
                  temperature=temperature, density=density_callback.address)

# ---------------------------setting initial field-------------------------------
@cfunc(types.field_loop_callback)
def setField_callback(ind, r, E, B, data_double, data_int):
    E[0] = field_amplitude*np.sin(2*np.pi*r[0]/(xmax-xmin))

sim.field_loop(handler=setField_callback.address)


# =================================OUTPUT========================================
fig, axs = plt.subplots(2, constrained_layout=True)

# -------------preparing output for electron distribution f(x, px)--------------
xpx_dist = np.zeros((64, 128), dtype=np.double)
pxLim = 5*np.sqrt(temperature*electron_mass)
inv_dx_dpx = (xpx_dist.shape[1]/(xmax-xmin))*(xpx_dist.shape[0]/(2*pxLim))

@cfunc(types.particle_loop_callback)
def xpx_callback(r, p, w, id, data_double, data_int):
    ix = int(xpx_dist.shape[1]*(r[0]-xmin)/(xmax-xmin))
    iy = int(xpx_dist.shape[0]*0.5*(1+p[0]/pxLim))
    data = carray(data_double, xpx_dist.shape, dtype=np.double)
    if iy >= 0 and iy < xpx_dist.shape[0]:
        data[iy, ix] += w[0]*inv_dx_dpx/(3*density/pxLim)

axs[0].set_title('$\partial N / \partial x \partial p_x$ (s g$^{-1}$cm$^{-2}$)')
axs[0].set(ylabel='$p_x$ (cm g/s)')
axs[0].xaxis.set_ticklabels([])
plot0 = axs[0].imshow(xpx_dist, vmin=0, vmax=1,
                      extent=[xmin, xmax, -pxLim, pxLim], interpolation='none',
                      aspect='auto', cmap='YlOrBr')
fig.colorbar(plot0, ax=axs[0], location='right')

def plot_xpx():
    xpx_dist.fill(0)
    sim.particle_loop(name='electron', handler=xpx_callback.address,
                      data_double=pipic.addressof(xpx_dist))
    plot0.set_data(xpx_dist)

# -------------------------preparing output of Ex(x)-----------------------------
Ex = np.zeros((32,), dtype=np.double)

@cfunc(types.it2r_callback)
def Ex_it2r(it, r, data_double, data_int):
    r[0] = xmin+(it[0]+0.5)*(xmax-xmin)/Ex.shape[0]

@cfunc(types.field2data_callback)
def get_Ex(it, r, E, B, data_double, data_int):
    data_double[it[0]] = E[0]

axs[1].set_xlim([xmin, xmax])
axs[1].set_ylim([-field_amplitude, field_amplitude])
axs[1].set(xlabel='$x$ (cm)', ylabel='$E_x$ (cgs units)')
x_axis = np.linspace(xmin, xmax, Ex.shape[0])
plot_Ex_, = axs[1].plot(x_axis, Ex)

def plot_Ex():
    sim.custom_field_loop(number_of_iterations=Ex.shape[0], it2r=Ex_it2r.address,
                       field2data=get_Ex.address,data_double=pipic.addressof(Ex))
    plot_Ex_.set_ydata(Ex)


# ===============================SIMULATION======================================
output_folder = 'basic_example_output'
if not os.path.exists(output_folder):
    os.makedirs(output_folder)
time_start = time.time()
for i in range(32):
    sim.advance(time_step=time_step, number_of_iterations=2)
    plot_xpx()
    plot_Ex()
    fig.savefig(output_folder + '/im' + str(i) + '.png')
    if i == 25:
        fig.savefig(output_folder + '/fig2.pdf')
    print(i, '/', 32)
\end{minted}

\begin{figure} 
\centering\includegraphics[width=0.6\columnwidth, scale=1]{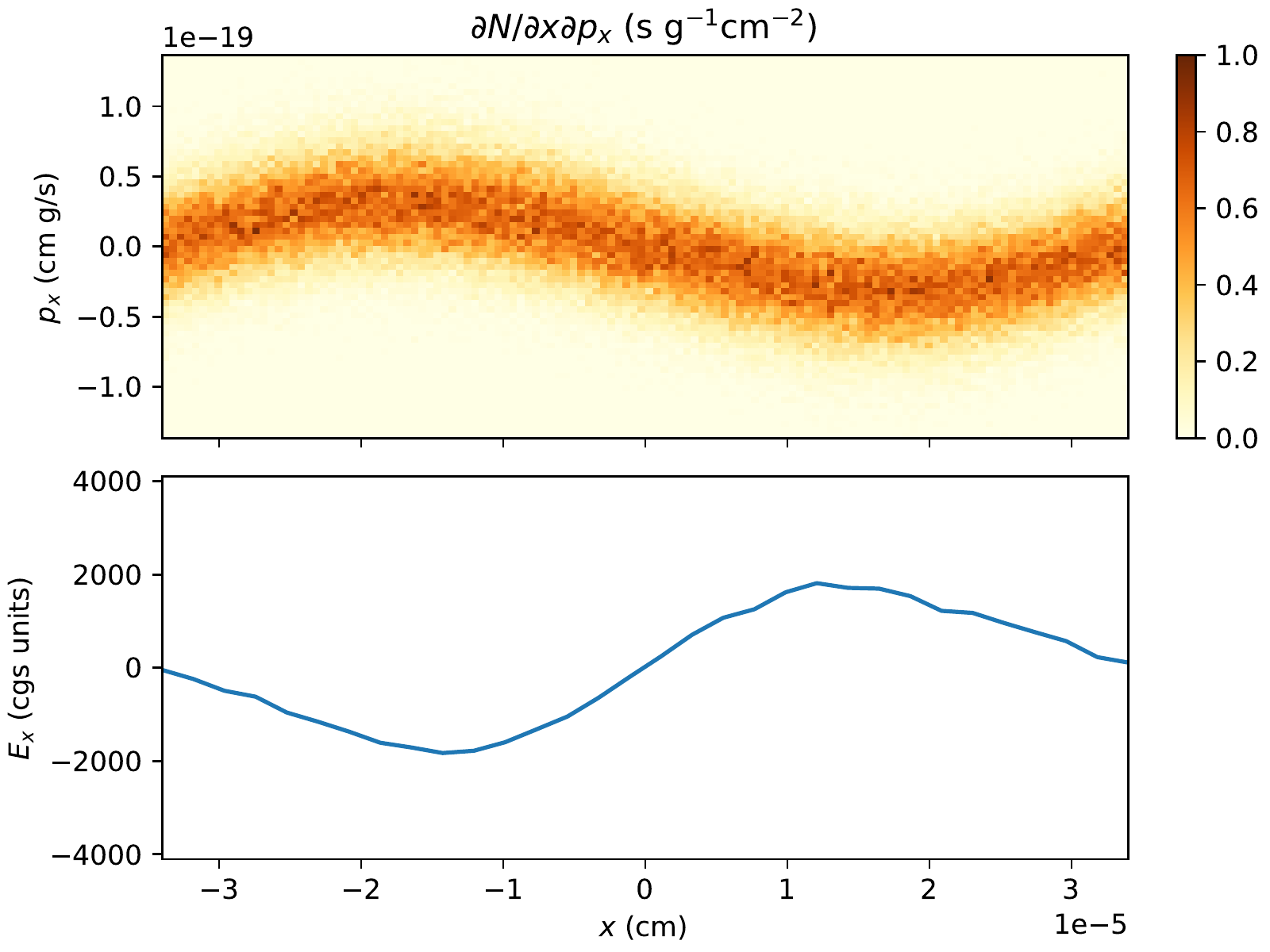}
\caption{The 25-th image created by the example presented in subsection \ref{sec_example}.}
\label{im25}
\end{figure}

The code produces a sequence of images. The 25-th image is shown in fig.~\ref{im25}. When choosing solver in line {21} it is possible to select {\texttt{fourier\_boris}, \texttt{ec} or \texttt{ec2}.}

\subsection{{Landau damping}}
\begin{figure} 
\centering\includegraphics[width=1.0\columnwidth, scale=1]{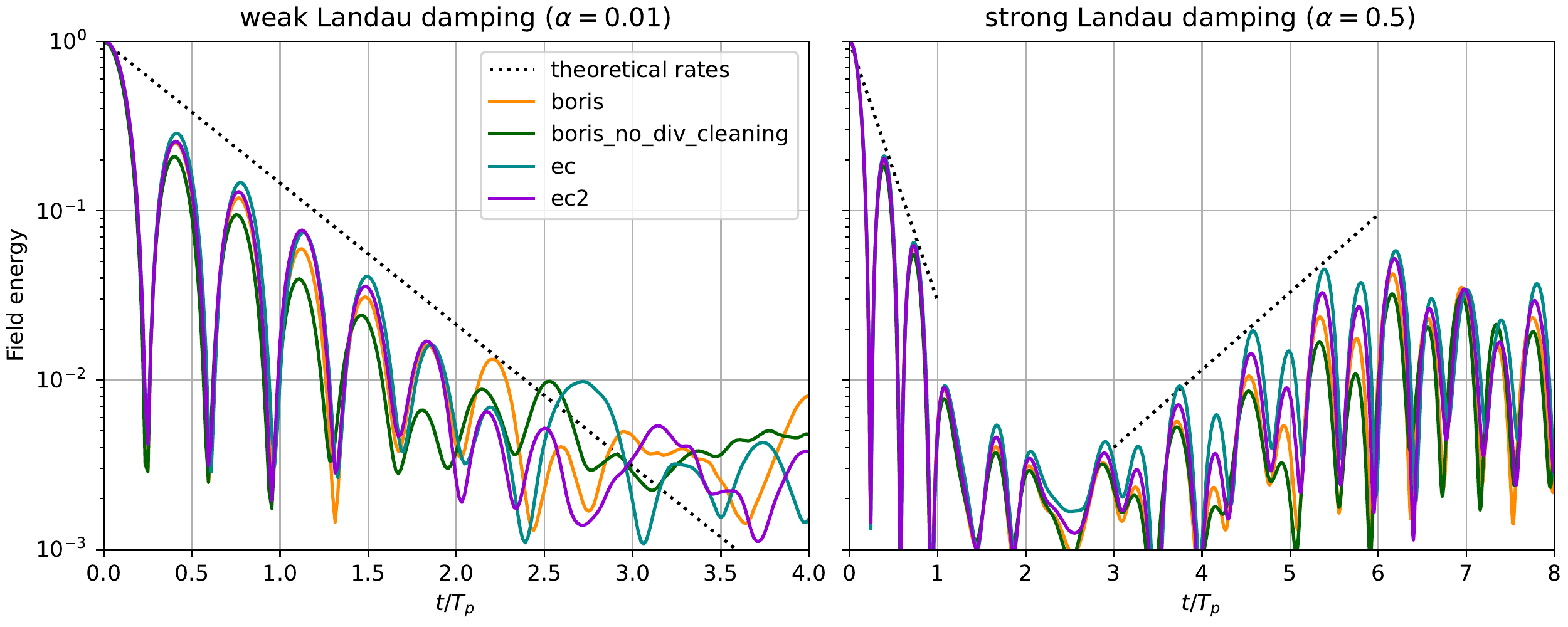}
\caption{{Verification of the developed solvers based on the simulation of Landau damping in cases described in Ref.~\cite{cheng.jcp.1976}. Label "boris\_no\_div\_cleaning" stands for Boris pusher with disabled divergence cleaning. The plots show temporal evolution of the energy of electric field normalized to the initial value. The slopes of theoretical rates (dotted lines) are taken from Ref.~\cite{cheng.jcp.1976}}}
\label{fig_landau_damping}
\end{figure}

{In this subsection we verify the correctness of all the implemented solvers by simulating Landau damping in cases of weak and strong perturbation. We consider parameters of the standard cases described in Ref.~\cite{cheng.jcp.1976}. We consider non-relativistic temperature $T = (2/3)10^{-6}mc^2$ and set the initial plasma density to $N(x) = 1 + \alpha \sin\left(2\pi x/(\texttt{XMax} - \texttt{XMin})\right)$, where the limits of the simulation region are $\texttt{XMax} = - \texttt{XMin} = 2\pi \lambda_D$ with $\lambda_D$ being the Debye length. The initial electric field is defined from the assumption of uniform positive background of ions. We use one-dimensional computational region with $N_x = 16$ cells and $10^4$ particles per cell. The time step is $\Delta t = T_p/64$. The results of the simulations and their comparison with theoretical rates for $\alpha = 0.01$ and $\alpha = 0.5$ are presented in fig.~\ref{fig_landau_damping}. Variation of computational parameters does not show any notable benefit of using energy conserving method in this problem. This is natural because the time and space steps needed to resolve the physics in question is much less the plasma period and Debye length. In this case the conventional Boris method is known to be stable and the numerical heating due to minor violation of energy conservation is happen to be negligible as compared to the scales of interest.}

\subsection{{Relativistic two-stream instability}}
\begin{figure} 
\centering\includegraphics[width=0.7\columnwidth, scale=1]{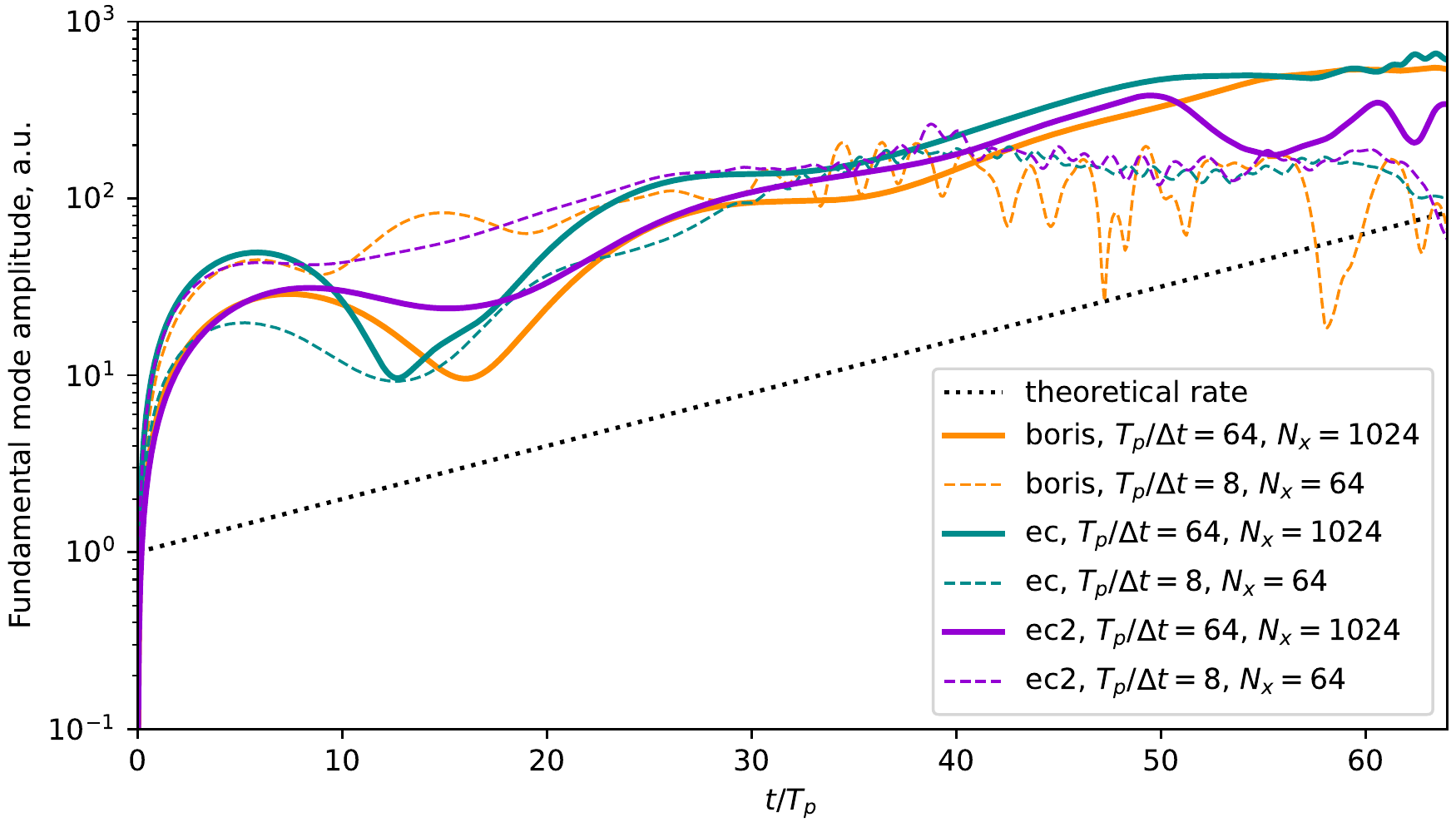}
\caption{{Verification of the developed solvers in relativistic case based on the simulation of two-stream instability.}}
\label{fig_tsi}
\end{figure}

{In order to verify the correctness of the solvers in relativistic case we simulate the relativistic two-stream instability and compare the results with the rate given in, e.g. Ref.~\cite{shalaby.aj.2017}. We consider two counter-propagating streams of electrons with equal density $N = 1/2$ and gamma factor being equal to $\gamma_s = 10$ for both streams. The initial momentum is modified by a factor $1 \pm 0.01\sin\left(2\pi x/(\texttt{XMax} - \texttt{XMin})\right)$ to set a more stable seed for the instability. Following the theory consideration presented in \cite{shalaby.aj.2017} the limits of computational regions $\texttt{XMax} - \texttt{XMin} = \lambda_0 \approx 1.6329 \gamma_s^{2/3} v_s T_p$ are set to include a single mode with maximal growth rate $\Gamma \approx 0.011 (2\pi/T_p)$ ($v_s$ is the velocity that corresponds to gamma factor $\gamma_s$). We again use $10^4$ particles per cell, but consider two options for the time and space step, the latter being defined via the number of grid points $\Delta x = (\texttt{XMax} - \texttt{XMin})/N_x$: $\Delta t = T_p/64$, $N_x = 1024$ (high resolution) and $\Delta t = T_p/8$, $N_x = 64$ (low resolution). For comparison we plot the amplitude of the fundamental mode as a function of time in logarithmic scale together with a line with the slope defined by the theoretical value of the growth rate $\Gamma$. The results for different solvers are compared in fig.~\ref{fig_tsi}. The capabilities of all the solvers to deal with the case of low resolution are similar. The only minor deviation worth mentioning is the rapid field amplitude oscillations observed for the "boris" solver at the late, highly non-linear stage of the simulated process.}

\section{Basic properties}
\label{sec_properties}

\subsection{The limit of low resolution}
\label{sec_limit}

Even thought the method warrants energy conservation by construction, we need to verify that its use is not precluded by numerical artefacts that do not violate this conservation law. For example, preserving energy per se does not contradict to the hypothetical process of transferring kinetic energy of thermal particle motion to electromagnetic field of plasma oscillations, i.e. to the process of spontaneous generation of plasma oscillations. (Clearly, such a behavior would contradict to the second law of thermodynamics; any plasma oscillations should in opposite decay in line with Landau damping.) To assess the presence of this and potentially other numerical artefacts in case of low resolution we perform a more extensive study of plasma oscillations as probably one of the most indicative plasma processes.

\begin{figure} 
\centering\includegraphics[width=1.0\columnwidth, scale=1]{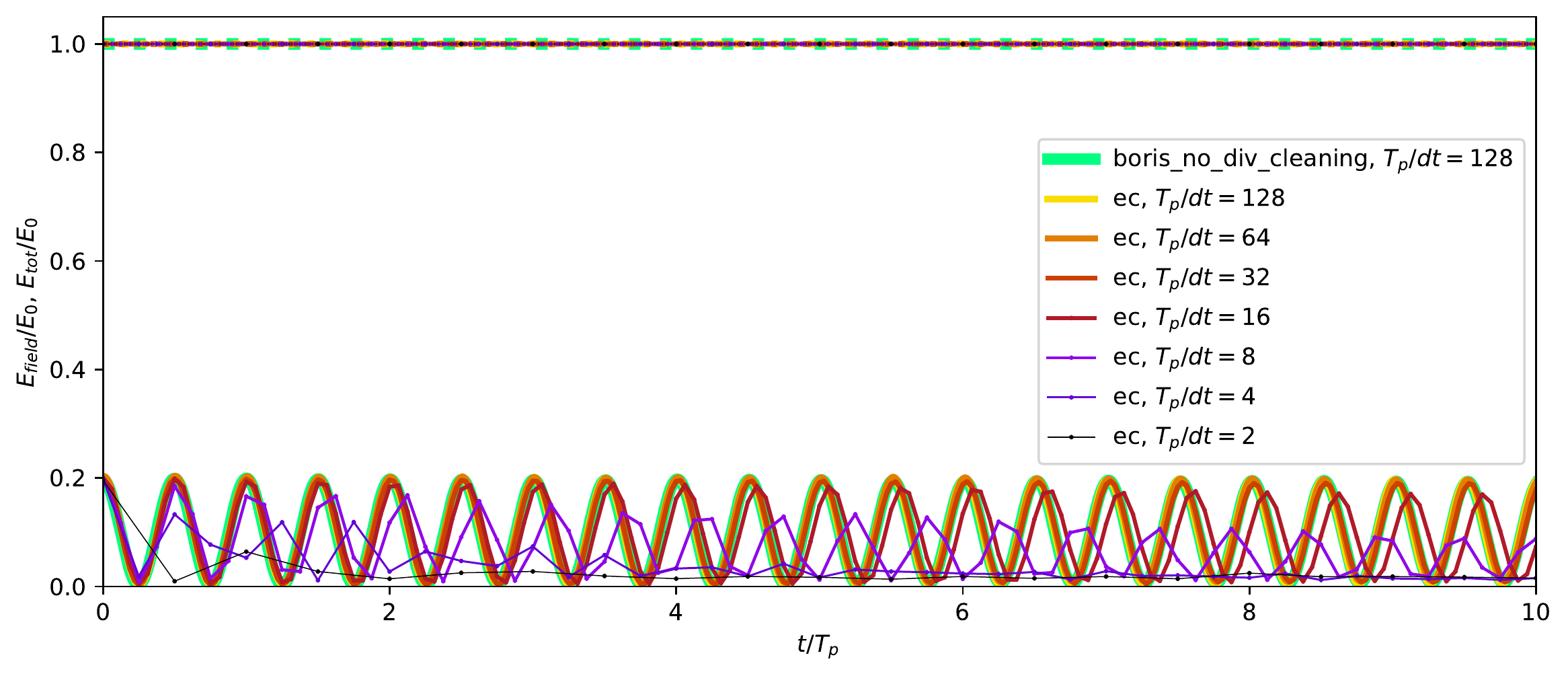}
\caption{{Analysis of the effect of low space and time resolution on an oscillating cold plasma simulations carried out with the energy conserving solver with first order accuracy (labeled "ec"). The plot shows the dependencies of electromagnetic $E_{field}$ (solid curves) and total $E_{tot}$ (dotted curves) energy on time $t$ during 10 plasma periods $T_p$ for difference values of time step $dt$. The values of energy are normalized to the initial energy $E_0$. The results are compared with high-resolution simulation carried out with Boris pusher coupled with the Fourier-based solver for Maxwell's equations (see details in the text) with disabled divergence cleaning (labeled as "boris\_no\_div\_cleaning").}}
\label{en_cons}
\end{figure}

We consider a simulation of 10 plasma periods for a setup described in Sec.~\ref{sec_example} with {\texttt{XMin, XMax = -612.4*DebyeLength, 612.4*DebyeLength}} represented by \texttt{nx = 32} cells, using 100 particles per cell and varied time step (see Appendix.~\ref{source_oscillations}). To test the {applicability limits of the developed solvers} we intentionally consider low resolution for both time and space. The results for different time resolution are presented in fig.~\ref{en_cons}. 

In case of using the energy-conserving method (labeled "ec") the total energy is preserved to within machine accuracy (deviation is below $10^{-11}$) independently {of} the time step. {As a benchmark we show the result of a simulation with high temporal resolution using the standard Boris pusher in combination with spectral field solver with disabled divergence cleaning. As one can see, for high resolution the results of both solvers are in a good agreement.}

Let us consider how the results change when resolution becomes low. We can see that in the limit of low resolution the presented energy-conserving method doesn't cause any more severe deviations than increased rate of decay and frequency altering, both effects being fairly minor until about 16 steps per plasma period. In this case, the increased rate of decay becomes significant for the resolution of 8 steps per plasma period, while even for 4 steps per period the effect of plasma oscillations is qualitatively reproduced. In case of 2 steps per plasma period the oscillations decay almost immediately, while still not causing any clearly destructive nonphysical plasma states. 

These results indicate that {the "ec" method can be found capable of performing simulations with a fairly large time step without causing any destructive numerical artefacts. This means that} even if the time step is insufficient for resolving plasma oscillations in some regions of high plasma density, these regions are expected to remain physically stable. The latter can be useful if the physics of interest doesn't require good resolution of many plasma oscillations, or such oscillations are resolved using reduced time step during a limited part of simulation.

\begin{figure} 
\centering\includegraphics[width=1.0\columnwidth, scale=1]{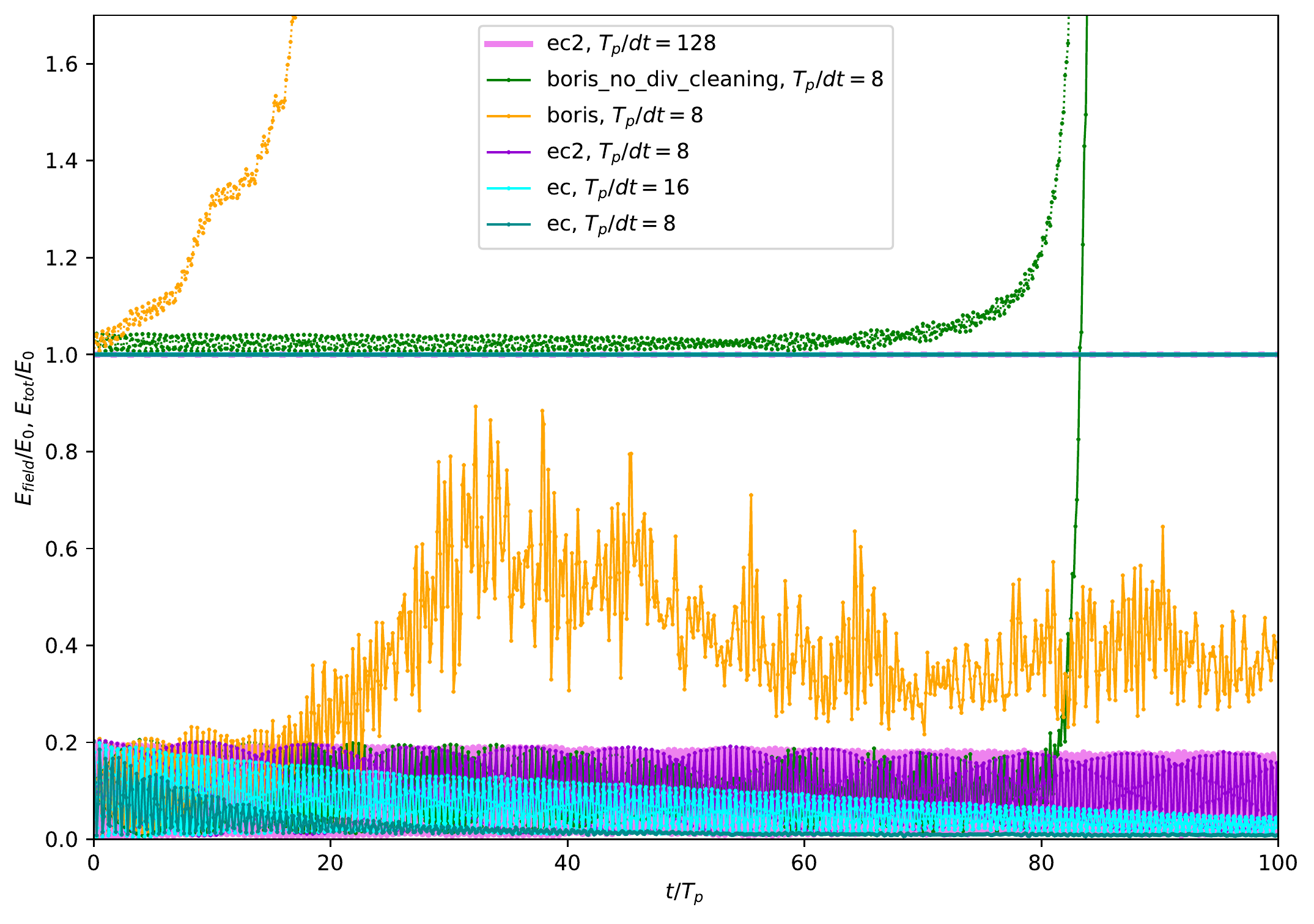}
\caption{{The comparison of the capabilities of different solvers shown with the same data as in fig.~\ref{en_cons} but for a time interval of $100 T_p$ and a sufficiently low space and time resolution to make visible the limitations. The case labeled "boris" stands for the Boris pusher with enabled divergence cleaning.}}
\label{en_cons1}
\end{figure}

{To identify the applicability limitations of all the developed solvers we extend the duration of simulation to 100 plasma oscillations and consider the results for a sufficiently low resolution to cause notable deviation from the exact solution. The results are presented in fig.~\ref{en_cons1}. The most accurate computations are achieved with "ec2" solver and thus in this case we use its result with high resolution ($T_p/dt = 128$) as a benchmark. One can see that the values obtained with the Boris pusher without divergence cleaning (labeled "boris\_no\_div\_cleaning") and $T_p/dt = 8$ fairly accurately follow the benchmark values up until $t \approx 80 T_p$ and then run into an instability. This instability is caused by accelerated accumulation of errors due to inaccurate current deposition and can be suppressed by divergence cleaning, which is well demonstrated by the result obtained with the same time step by the Boris pushed with enabled (by default) divergence cleaning (labeled "boris"). Nevertheless, one can see that the stability of the "boris" solver is achieved at the cost of notable continuous numerical heating that completely alters the structure of the field energy oscillations starting from about $t = 20 T_p$. Note, however, that such a rapid numerical heating is caused by an extremely low temporal and spatial resolution $\Delta x \approx 30 \lambda_D$, where $\lambda_D$ is the Debye length.}

{From fig.~\ref{en_cons1} we can also see that "ec2" solver provides a remarkably accurate result, which almost perfectly agrees with the benchmark (the deviation is rather caused by aliasing related to the fact of having just 4 points per each oscillation of field energy). At least the results is significantly more accurate than that obtained with "ec" solver with factor 2 smaller time step. Given that "ec2" involves two loops over particles per iteration, we can clearly see the benefit of achieving the second order accuracy.}

\subsection{The cost of using low resolution}
\label{sec_cost}

The results presented in the previous subsection indicate that in the limit of low resolution the use of energy-conserving method leads to an increased rate of decay for plasma oscillations. It is interesting to consider this effect in more detail. This is to reveal potentially less prominent effects that constitute the price we pay for using low resolution. For this purpose in this subsection we consider how a low resolution affects the particle distribution in the phase space.

Since in $\pi$-PIC the particles are stored and processed in a cell-by-cell order, the effect of order mentioned in the end of Sec.~\ref{sec_coupling} can clearly play a significant role. In case of single loop over all particles, the particles that arrive to a given cell are processed either before or after all other particles in this cell. Since the chance of migration is correlated with energy, this clearly introduces a bias. That is why when energy-conserving scheme is used in $\pi$-PIC the particles are shuffled in each cell (before each iteration) and processed in two halves: first we make a loop over all cells processing a half of particles and then make another loop over all cells processing the remaining particles. To see the effect of such shuffling, we present a separate result for simulation without it.

\begin{figure} 
\centering\includegraphics[width=0.7\columnwidth, scale=1]{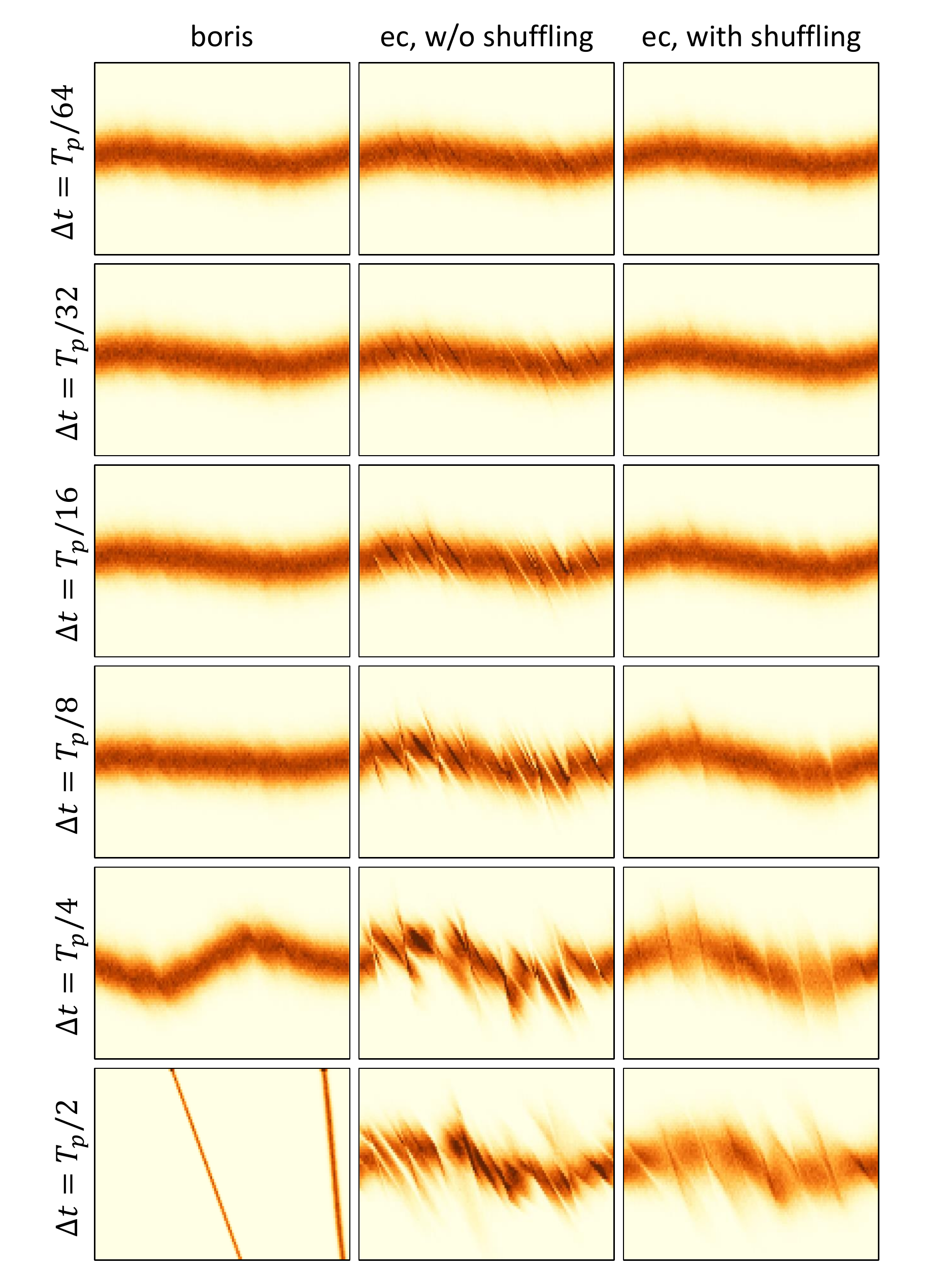}
\caption{Distributions of particles in the space of $x$-$p_x$ (for the description of axes and the color scheme see fig.~\ref{im25}) for different setups: Boris pusher with spectral field solver (left column), energy-conserving method without shuffling (central column), and energy-conserving method with shuffling (right column).}
\label{effect_of_order}
\end{figure}

To perform analysis we repeat the simulation of a single plasma oscillation described in sec.~\ref{sec_example} with \texttt{nx=8}. In fig.~\ref{effect_of_order} we show the phase space $x$-$p_x$ described in fig.~\ref{im25} for different time step values and for different setups: Boris pusher, energy-conserving method without shuffling and energy-conserving method with shuffling. We can see that shuffling indeed significantly suppresses the effect of order and overall improves the accuracy of simulation in the limit of low resolution. In case of low resolution the energy-conserving method with shuffling shows a rather reasonable distribution that doesn't seem to deviate dramatically from the thermal distribution, which one could observe after the decay of plasma oscillations.

\section{{Verification} and capabilities}
\label{sec_verification}

In this section we pursue two goals. First, we {verify} the correctness of the $\pi$-PIC code. Second, we show that in a rather generic case the proposed energy-conserving method indeed makes the code capable of performing simulations with time and space steps notably larger than that needed to resolve plasma period and Debye length, respectively.

The {verification} and comparison can be more indicative if the simulated process involves complex relativistic plasma dynamics. In this case exact analytical results that provide complex data for comparison are rarely achievable. One example is the theory for relativistic radiation reflection from dense plasmas, referred to as Relativistic Electronic Spring (RES) \cite{gonoskov.pre.2011, gonoskov.pop.2018}. The RES theory provides a way to compute the field structure of the electromagnetic pulse after reflection from a plasma surface with arbitrary density distribution. The results become increasingly more accurate for higher intensity of the incident pulse so that highly relativistic plasma dynamics is induced. In case of non-zero incidence angle this process can still be simulated in 1D geometry using a moving reference frame (in $\pi$-PIC this can be arranged by setting initial moment for all particles). Nevertheless, we intentionally consider the process in laboratory reference frame to make testing more comprehensive. We choose to perform {verification} and comparison using 2D simulations (3D case can be considered by setting \texttt{nz}$\neq 1$ and modifying the shape of the pulse along $z$). 

We consider a short circularly polarized pulse with the wavelength of $\lambda = 1 \: \mu$m impinging on a plasma layer with the incidence angle of $\theta = \pi/3$. The layer has smooth density drops within $1 \: \mu$m and the shape of $\propto \sin^2$. The bulk density is $N_0 = 100n_{cr}$, where $n_{cr} = \pi m c^2/e^2\lambda^2$ is the plasma critical density. The peak field amplitude of the pulse is $100 a_0$, where $a_0 = 2\pi m c^2 / \lambda |e|$ is the relativistic field amplitude. We perform simulations for different time and space steps keeping $2 c\Delta t = \Delta x = \Delta y$. The initial plasma temperature is chosen so that the Debye length is set to $\lambda_D = \lambda/640$. {In all simulations we use 30 particles per cell for the inner par of the plasma region.} Other details can be read from the listing of the Python code presented in Appendix.~\ref{source_verification}. Note that this setup is related to experiments on the interaction of intense laser pulses with solid targets, which is motivated by various applications including high harmonic generation and particle acceleration. 

\begin{figure} 
\centering\includegraphics[width=0.7\columnwidth, scale=1]{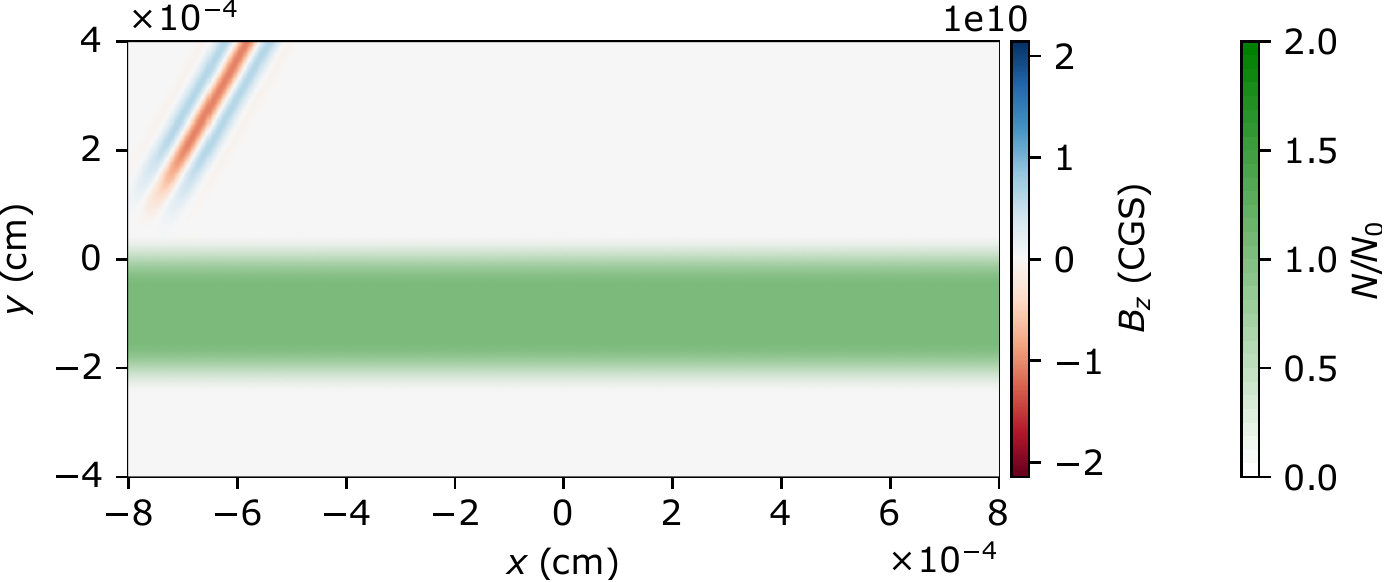}
\caption{Field $B_z$ component and plasma density $N$ normalized to $N_0 = 100 n_{cr}$ in the initial moment of simulations. The pulse propagates towards point $x = y = 0$.}
\label{im0}
\end{figure}

\begin{figure} 
\centering\includegraphics[width=1.0\columnwidth, scale=1]{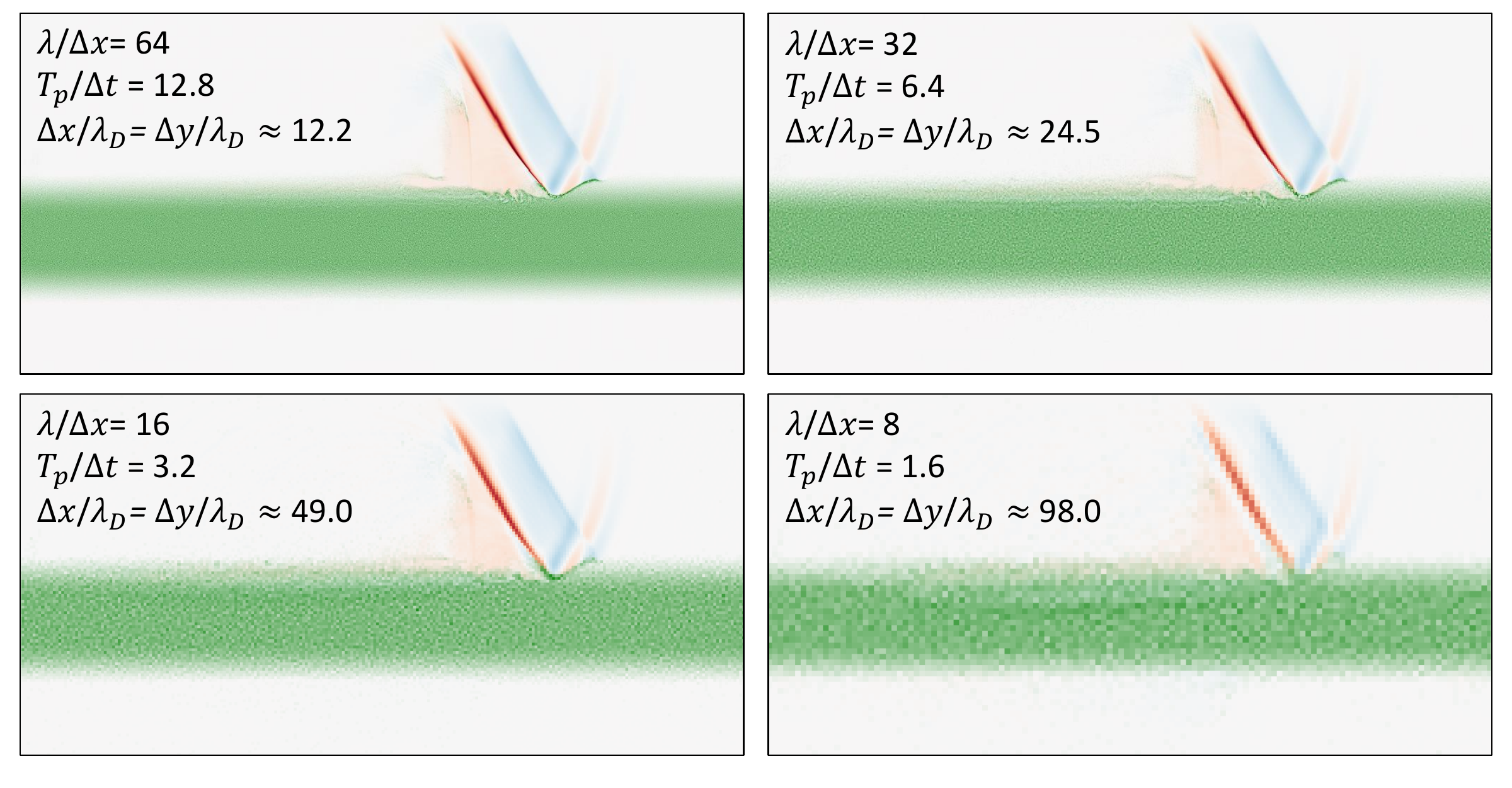}
\caption{{Results} of simulations for different time and space steps. For the description of axes and color maps see fig.~\ref{im0}.}
\label{im21}
\end{figure}

\begin{figure} 
\centering\includegraphics[width=0.7\columnwidth, scale=1]{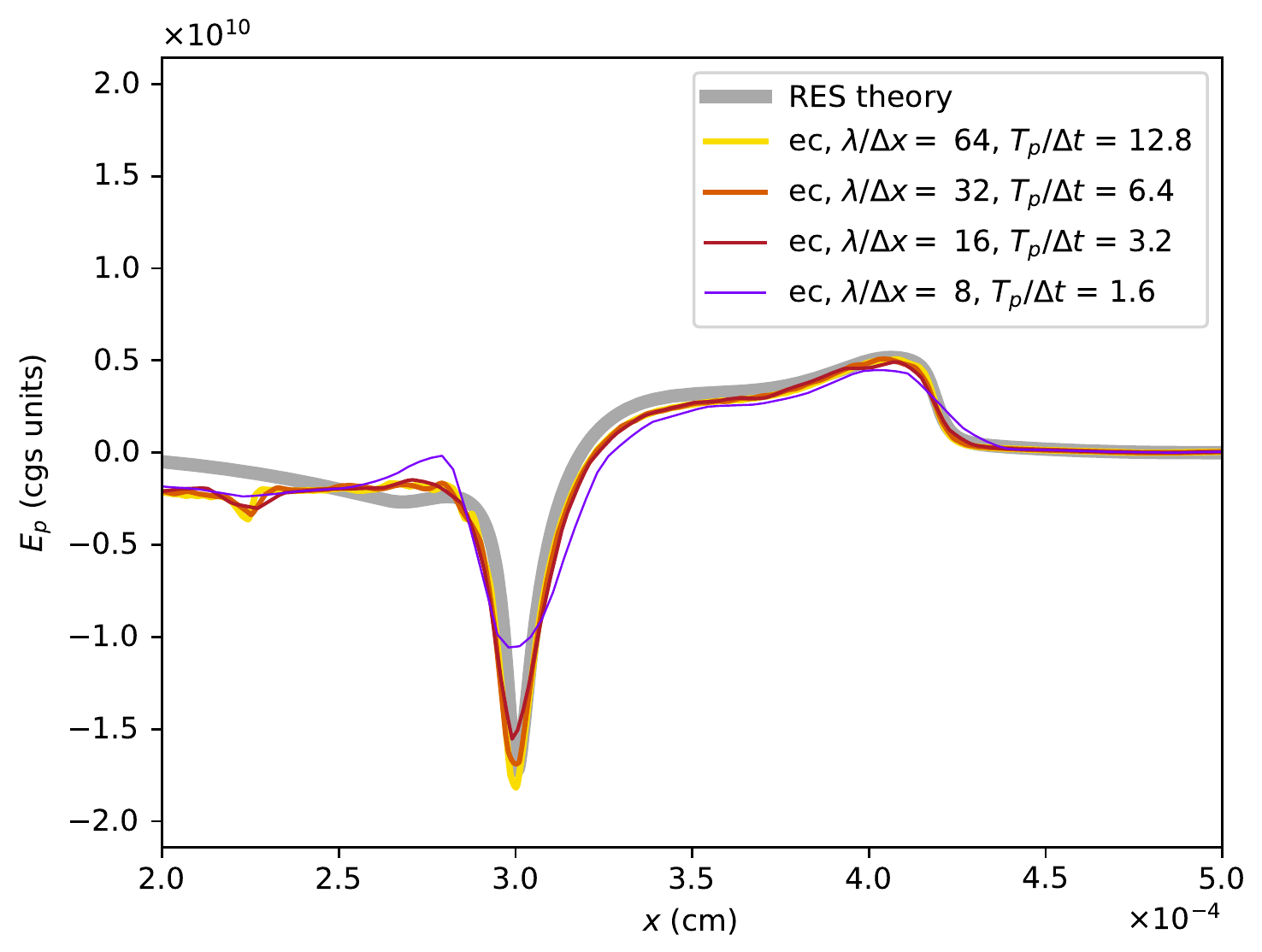}
\caption{Comparison of the transverse electric field obtained using the RES theory and simulations with different time and space steps.}
\label{res_comparison}
\end{figure}

In fig.~\ref{im0} we show the initial state of the field and plasma. The pulse, shown via $B_z$ field component, propagates toward point $x = y = 0$. The state of the system at the time instance $t \approx 3.5 \times 10^{-14}$ s is shown in fig.~\ref{im21} for all considered cases. One can see the outgoing pulse that can be benchmarked against the RES results. In fig.~\ref{res_comparison} we compare the transverse field along the line $x=y \tan\theta$ at $t \approx 3.5 \times 10^{-14}$ s. 

From fig.~\ref{res_comparison} we can see that numerical simulations result in shapes that are close to what obtained with the RES theory. This indicates that the code is likely to work correctly. {Similar results are obtained for all the developed solvers. Certainly the computation based on the RES theory provides only an approximate result that is useful for verifying the correctness of implementation, but not for comparing the capabilities of the solvers. We should not expect the exact agreement with the RES theory because of many reasons including limited transverse size and limited amplitude of the pulse (the results obtained with the RES theory become increasingly accurate with the increase of incident radiation intensity).} We can also see that the results remain fairly accurate even in case the time step is comparable with the plasma period $T_p = \lambda/10c$ for the density inside the target. Note that in the case of $T_p/\Delta t = 1.6$, it is likely that the low spatial resolution of the wavelength $\lambda$ affects the accuracy of the simulation. 

\begin{figure} 
\centering\includegraphics[width=0.7\columnwidth, scale=1]{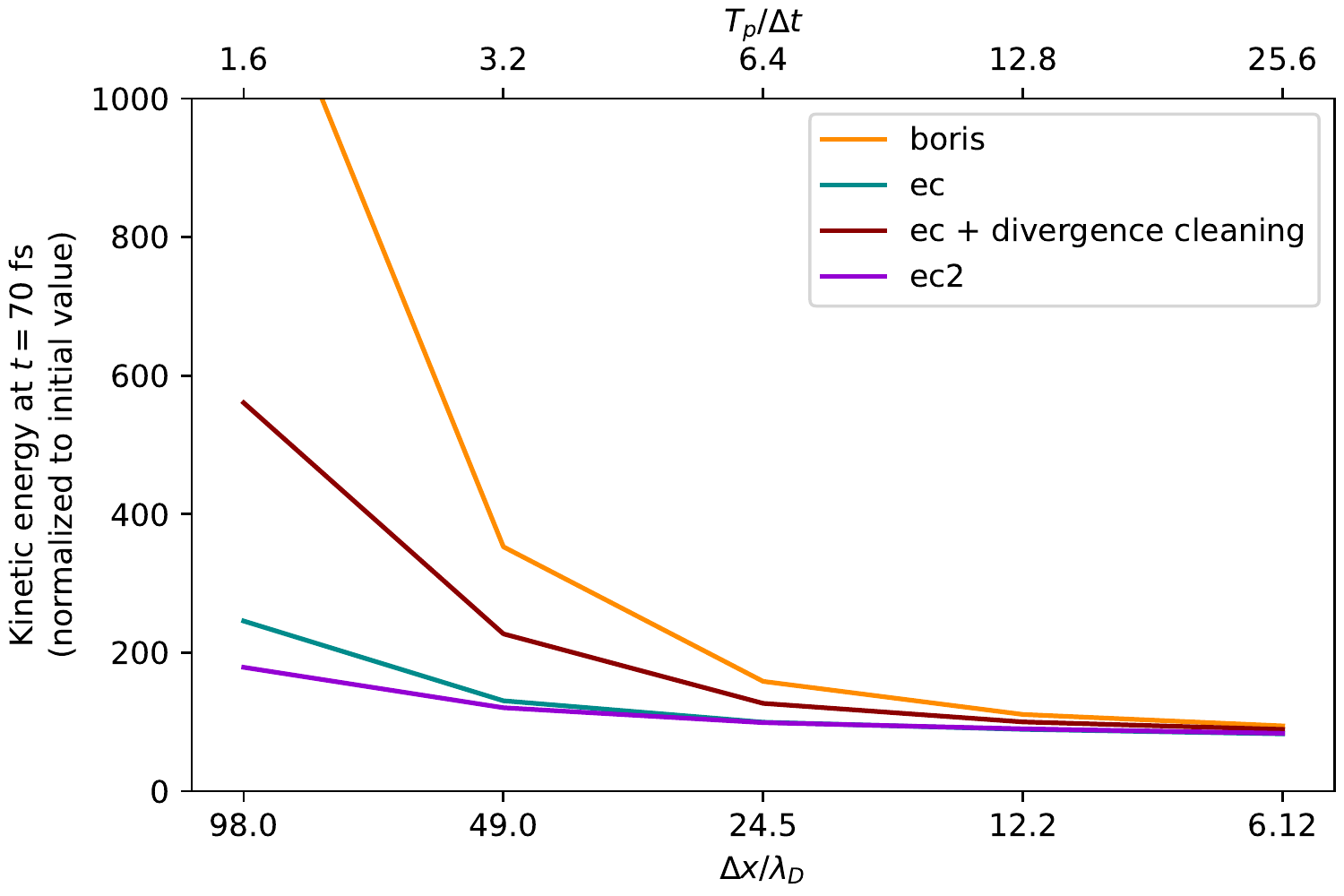}
\caption{{Comparison of the capabilities of different solvers for simulating plasma heating in the process of intense electromagnetic pulse interaction with overdense plasma. The total kinetic energy after the interaction (at $t = 70$ fs) is normalized to the initial value and plotted as a function of numerical resolution for different solvers.}}
\label{fig_heating}
\end{figure}

{In the context of this numerical experiment one characteristic property of the proposed method is the absence of numerical heating, which is a common consequence of imperfect energy conservation in case of insufficient time and space resolution for conventional schemes. To assess a potential benefit we consider the process of plasma heating by the incident electromagnetic pulse. In fig.~\ref{fig_heating} we plot the kinetic energy of electrons after the interaction at $t = 70$ fs normalized to the initial value at $t = 0$ fs as a function of resolution for different numerical schemes. As previously, the “boris” scheme here denotes a combination of using Boris pusher with CIC weighting in combination with the dispersion-free Fourier solver of Maxwell’s equations and divergence cleaning for charge conservation. The increased kinetic energy observed for the Boris pusher is related to numerical heating, while for energy conserving methods the increased energy at small resolution can be associated with overestimation of energy coupling which can naturally occur due to order-of-one-step deeper penetration of electromagnetic radiation behind a sharp radiation-plasma interface. One can see that as compared to the results of “boris” scheme, energy conserving schemes achieve similar accuracy levels with factor 2-4 larger space and time steps, which can be roughly translated to a speedup factor 8 -- 64 for 2D case and factor 16 -- 256 for 3D case. This indicates that the proposed method can be useful for long-term simulations that require absence of numerical heating. One possible application is the studies of laser-driven ion acceleration by the mechanism of Target Normal Sheath Acceleration \cite{macchi.rmp.2013}. The method or its improved versions can be also found useful for suppressing numerical instabilities in simulations of laser wake-field acceleration \cite{esarey.rmp.2009} in a boosted reference frame \cite{yu.jcp.2014}.}

\begin{figure} 
\centering\includegraphics[width=1.0\columnwidth, scale=1]{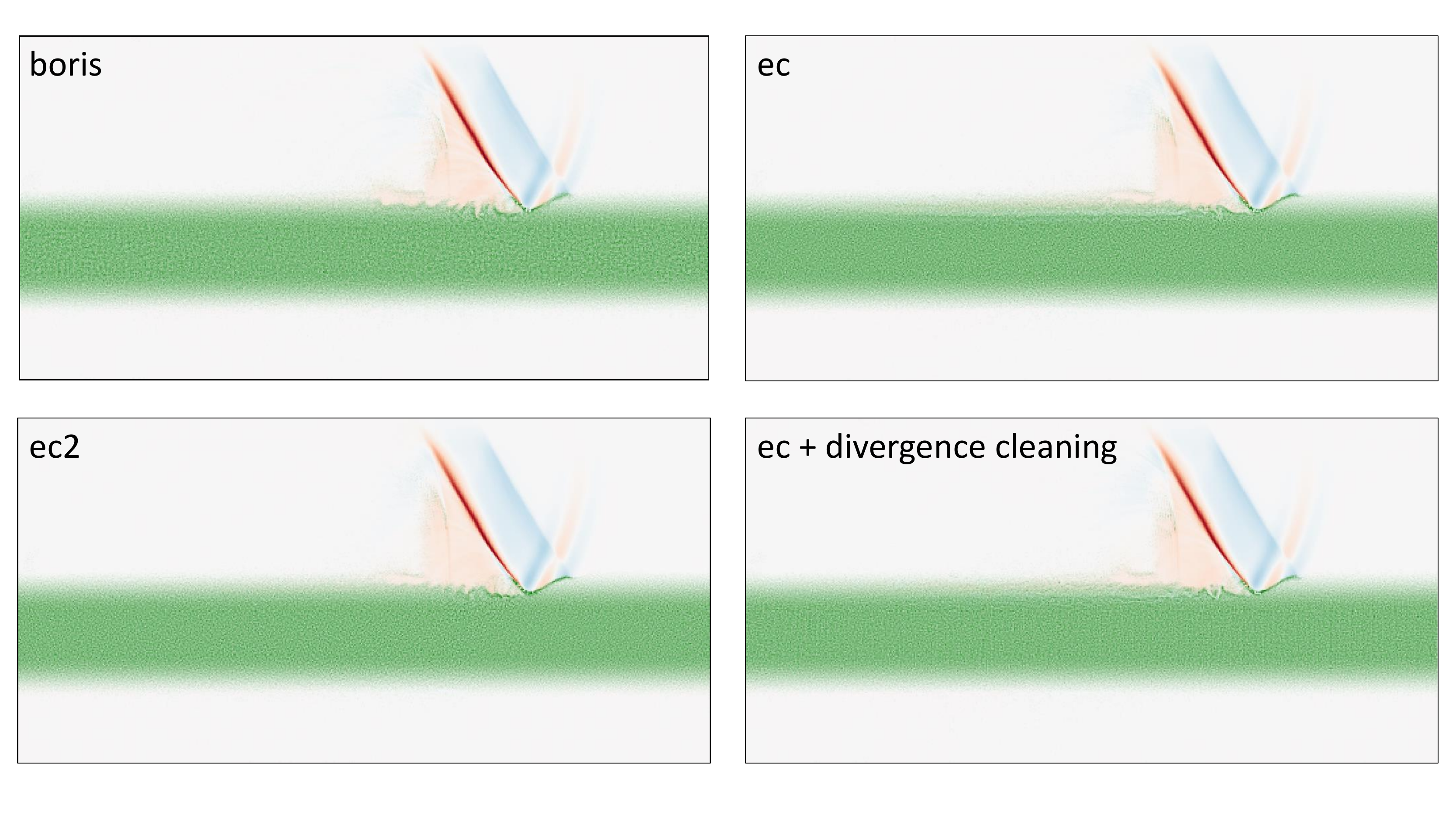}
\caption{{Comparison of the system state at $t = 35$ fs obtained with different solvers. The axes are the same as in fig.~\ref{im0}.}}
\label{fig_heating_comp}
\end{figure}

{Although the density distribution and field states shown in fig.~\ref{im21} look almost identical for all the solvers, let us make a more detailed consideration of the differences between the solvers. First, we can note that the combination of energy conserving solver with divergence cleaning (labeled “ec + divergence cleaning”) has no property of exact energy conservation, giving an origin for numerical heating, which however is slightly weaker than that in the case of “boris” solver. This indirectly indicates that in case of low resolution the energy conservation appears to be in a increasingly notable disagreement with charge conservation (otherwise divergence cleaning would not have any effect). To see this effect, we plot the state of system at $t = 35$ fs for different solvers in fig.~\ref{fig_heating_comp}. In case of “ec”  behind the sliding reflection point one can note a trace of rapid density drop, which does not oscillate and in the absence of moving ions can clearly be associated with accumulated deviation from the charge conservation. Indeed, this trace is suppressed outside interaction region in case of applying divergence cleaning to the “ec” solver. This indicates that imposing charge conservation within the proposed approach is likely to be extremely useful and can significantly improve accuracy. The second order energy conserving method (“ec2”) is likely to suffer from imperfect charge conservation but the higher accuracy seem to make it much weaker, so that it is not clearly visible in fig.~\ref{fig_heating_comp}.}

{Although the results indicate a potential usefulness of the proposed scheme, we should admit that we here do not consider many possible improvements to the standard scheme, including higher order form factors for particles, current smoothing and/or high harmonic filtering (see e.g. \cite{arber.ppcf.2015, fiuza.ppcf.2011}). Note however that in contrast to the presented method, the use of these improvements can imply an effective reduction of resolution, increased computational demands for larger number of distant memory calls, as well as a continuous energy dissipation and/or artificial non-local field modification. At the same time, as described above, there are several possible improvements to the proposed scheme, including imposed charge and momentum conservation. A more conclusive analysis requires further developments and studies with more elaborate measures of accuracy.}

\section{Conclusions}
\label{sec_conclusion}

In this paper we described a way to modify the standard relativistic PIC method so that the exact energy conservation is enforced, while the computational scheme remains explicit and involves similar amount of computations and memory calls. We derived governing principles and a basic algorithm for their implementation. We considered possible limitations and some ways for overcoming them. To facilitate further studies and use of the proposed method we developed and made publicly available the $\pi$-PIC code, which is relativistic energy-conserving PIC code based on spectral field solver \cite{pipic}. The performed tests indicate that the code and the proposed approach in general can be found useful for plasma simulations.  

\section{Acknowledgements}
The work is supported by the Swedish Research Council (Grants No. 2017-05148 and 2019-02376). {The author would like to thank Joel Magnusson for the help with $\pi$-PIC interfaces and deployment.}

\begin{appendices}

\section{{Python code used for Sec.~\ref{sec_limit}}}
\label{source_oscillations}

\begin{minted}[fontsize=\small,xleftmargin=20pt,linenos]{python}
import pipic
from pipic import consts, types
import matplotlib.pyplot as plt
import numpy as np
from numba import cfunc, carray
from pipic.consts import electron_mass, electron_charge, light_velocity


#===========================SIMULATION INITIALIZATION===========================
temperature = (2/3) * 1e-6 * electron_mass * light_velocity**2
density = 1e+18
debye_length = np.sqrt(temperature / (4*np.pi * density * electron_charge**2))
plasma_period = np.sqrt(np.pi * electron_mass / (density * electron_charge**2))
xmin, xmax = -612.4*debye_length, 612.4*debye_length
field_amplitude = 0.001 * 4*np.pi * (xmax - xmin) * electron_charge * density
nx = 32 # number of cells
iterations_period = 8
time_step = plasma_period/iterations_period
fig_label = 'ec2'

#---------------------setting solver and simulation region----------------------
if fig_label == 'boris':
    sim = pipic.init(solver='fourier_boris', nx=nx, xmin=xmin, xmax=xmax)
if fig_label == 'boris_no_div_cleaning':
    sim = pipic.init(solver='fourier_boris', nx=nx, xmin=xmin, xmax=xmax)
    sim.fourier_solver_settings(divergence_cleaning=False)
if fig_label == 'ec':
    sim = pipic.init(solver='ec', nx=nx, xmin=xmin, xmax=xmax)
if fig_label == 'ec2':
    sim = pipic.init(solver='ec2', nx=nx, xmin=xmin, xmax=xmax)

#------------------------------adding electrons---------------------------------
@cfunc(types.add_particles_callback)
def density_callback(r, data_double, data_int):# callback function
    return density # can be any function of coordinate r[0] 
sim.add_particles(name='electron', number=nx*100,
                 charge=-electron_charge, mass=electron_mass,
                 temperature=temperature, density=density_callback.address)

#---------------------------setting initial field-------------------------------
shift = np.pi/nx # a shift to mitigate aliasing

@cfunc(types.field_loop_callback)
def setField_callback(ind, r, E, B, data_double, data_int):
    E[0] = field_amplitude*np.sin(2*np.pi*r[0]/(xmax - xmin) + shift)

sim.field_loop(handler=setField_callback.address)


#=================================OUTPUT========================================
fig, axs = plt.subplots(2, constrained_layout=True)

#-------------preparing output for electron distrribution f(x, px)--------------
xpx_dist = np.zeros((64, 128), dtype=np.double)
pxLim = 5*np.sqrt(temperature * electron_mass)
inv_dx_dpx = (xpx_dist.shape[1]/(xmax - xmin))*(xpx_dist.shape[0]/(2*pxLim))

@cfunc(types.particle_loop_callback)
def xpx_callback(r, p, w, id, data_double, data_int):   
    ix = int(xpx_dist.shape[1]*(r[0] - xmin)/(xmax - xmin))
    iy = int(xpx_dist.shape[0]*0.5*(1 + p[0]/pxLim))
    data = carray(data_double, xpx_dist.shape, dtype=np.double)
    if iy >= 0 and iy < xpx_dist.shape[0]:
        data[iy, ix] += w[0]*inv_dx_dpx

axs[0].set_title('$\partial N / \partial x \partial p_x$ (s g$^{-1}$cm$^{-2}$)')
axs[0].set(ylabel='$p_x$ (cm g/s)')
axs[0].xaxis.set_ticklabels([])
plot0 = axs[0].imshow(xpx_dist, vmax=3*density/pxLim,
                      extent=[xmin,xmax,-pxLim,pxLim], interpolation='none',
                      aspect='auto', cmap='YlOrBr')
fig.colorbar(plot0, ax=axs[0], location='right')

def plot_xpx():
    xpx_dist.fill(0)
    sim.particle_loop(name='electron', handler=xpx_callback.address,
                      data_double=pipic.addressof(xpx_dist))
    plot0.set_data(xpx_dist)

#--------------------preparing output of kinetic energy-------------------------
mc2 = electron_mass * light_velocity**2
m2c2 = (electron_mass * light_velocity)**2
kineticEnergy = np.zeros((1, ), dtype=np.double)

@cfunc(types.particle_loop_callback)
def kineticEn_cb(r, p, w, id, data_double, data_int):   
    data_double[0] += w[0]*mc2*(np.sqrt(1+(p[0]**2+p[1]**2+p[2]**2)/m2c2)-1)

def getKineticEn(sim):
    kineticEnergy[0] = 0
    sim.particle_loop(name='electron', handler=kineticEn_cb.address,
                      data_double=pipic.addressof(kineticEnergy))
    return kineticEnergy[0]

#--------------------preparing output of field energy-------------------------
factor = ((sim.xmax-sim.xmin)/sim.nx)*((sim.ymax-sim.ymin)/sim.ny)* \
        ((sim.zmax-sim.zmin)/sim.nz)/(8*np.pi)
field_energy = np.zeros((1, ), dtype=np.double)

@cfunc(types.field_loop_callback)
def fieldEn_cb(ind, r, E, B, data_double, data_int):
    data_double[0] += factor*(E[0]**2+E[1]**2+E[2]**2+B[0]**2+B[1]**2+B[2]**2)

def getFieldEnergy(sim):
    field_energy[0] = 0
    sim.field_loop(handler=fieldEn_cb.address, 
                   data_double=pipic.addressof(field_energy))
    return field_energy[0]


#===============================SIMULATION======================================
n_plasma_periods = 100 # number of plasma oscillations to be simulated
experiment_duration = n_plasma_periods*plasma_period
n_iterations = n_plasma_periods*iterations_period
timeP = np.zeros((0, ), dtype=np.double)
Efield = np.zeros((0, ), dtype=np.double)
Etot = np.zeros((0, ), dtype=np.double)
E0 = 0
for i in range(n_plasma_periods*iterations_period):
    print(i, "/", n_plasma_periods*iterations_period)
    Ek = getKineticEn(sim)
    Ef = getFieldEnergy(sim)
    if i == 0:
        E0 = Ek+Ef  
    timeP = np.append(timeP, i*time_step/plasma_period)
    Efield = np.append(Efield, Ef/E0)
    Etot = np.append(Etot, (Ek+Ef)/E0)
    print('relative energy deviation = ', ((Ek+Ef)-E0)/E0)
    if Ef > 2*E0:
        break
    sim.advance(time_step=time_step)


#=====================SAVING SIMULATION RESULT TO FILE==========================
name = fig_label + '_' + str(iterations_period)
with open(name + '.npy', 'wb') as f:
    np.save(f, Etot)
    np.save(f, Efield)
    np.save(f, timeP)
\end{minted}

\section{{Python code used for Sec.~\ref{sec_verification}}}
\label{source_verification}

\begin{minted}[fontsize=\small,xleftmargin=20pt,linenos]{python}
import pipic
from pipic import consts, types
import matplotlib as mpl
import matplotlib.pyplot as plt
from mpl_toolkits.axes_grid1 import make_axes_locatable
import numpy as np
from numba import cfunc, carray, types as nbt
import os
from pipic.consts import electron_mass, electron_charge, light_velocity


#===========================SIMULATION INITIALIZATION===========================
wavelength = 1e-4
nx, xmin, xmax = 128, -8*wavelength, 8*wavelength
ny, ymin, ymax = 64, -4*wavelength, 4*wavelength
dx, dy = (xmax - xmin)/nx, (ymax - ymin)/ny
time_step = 0.5 * dx / light_velocity
figStride = int(nx/16)

@cfunc(nbt.double(nbt.double,nbt.double,nbt.double)) #auxiliary function
def cos2shape(x, plateauSize, transitionSize):
    return (abs(x) < plateauSize/2 + transitionSize) * \
           (1 - (abs(x) > plateauSize/2) *
           np.sin(0.5*np.pi*(abs(x) - plateauSize/2)/transitionSize)**2)

#---------------------setting solver and simulation region----------------------
sim=pipic.init(solver='ec2',nx=nx,ny=ny,xmin=xmin,xmax=xmax,ymin=ymin,ymax=ymax)

#---------------------------setting field of the pulse--------------------------
amplitude_a0 = 100
omega = 2 * np.pi * light_velocity / wavelength
a0 = electron_mass * light_velocity * omega / (-electron_charge)
fieldAmplitude = amplitude_a0*a0
arrivalDelay = 6*2*np.pi/omega
incidenceAngle = np.pi/3
# setup for a basic field generator/absorber
boundarySize = wavelength
supRate = 1 - np.exp(np.log(0.01) / (wavelength / (light_velocity * time_step)))

@cfunc(nbt.double(nbt.double, nbt.double)) # longitudinal shape
def pulse(eta, phi):
    return (abs(eta-2*np.pi)< 2*np.pi)*np.sin(eta/4)**3* \
           (np.cos(eta/4)*np.sin(eta+phi)+np.sin(eta/4)*np.cos(eta+phi))

@cfunc(types.field_loop_callback)
def field_callback(ind, r, E, B, data_double, data_int):
    if data_int[0] == 0 or ymax - r[1] < boundarySize:
        # computing longitudinal and transverse coordinates:
        lCoord = np.sin(incidenceAngle) * r[0] - np.cos(incidenceAngle) * r[1] + \
                 + (arrivalDelay - data_int[0]*time_step) * light_velocity
        tCoord = np.cos(incidenceAngle)*r[0] + np.sin(incidenceAngle)*r[1] 
        tShape = cos2shape(tCoord, 3*wavelength, 2*wavelength)
        # computing both field components of the CP pulse
        f_1=fieldAmplitude*tShape*pulse(2*np.pi*lCoord/wavelength+4*np.pi,np.pi)
        f_2=fieldAmplitude*tShape*pulse(2*np.pi*lCoord/wavelength+4*np.pi,np.pi/2)
        Ex = np.cos(incidenceAngle)*f_1
        Ey = np.sin(incidenceAngle)*f_1
        Bz = f_1
        Bx = -np.cos(incidenceAngle)*f_2
        By = -np.sin(incidenceAngle)*f_2
        Ez = f_2
        if data_int[0] == 0: # setting initial field
            E[0], E[1], E[2], B[0], B[1], B[2] = Ex, Ey, Ez, Bx, By, Bz
        else: # soft reduction of the field difference (pulse generation)
            rate = supRate*cos2shape(ymax - r[1], 0, boundarySize)
            E[0] = (1 - rate)*E[0] + rate*Ex
            E[1] = (1 - rate)*E[1] + rate*Ey
            E[2] = (1 - rate)*E[2] + rate*Ez
            B[0] = (1 - rate)*B[0] + rate*Bx
            B[1] = (1 - rate)*B[1] + rate*By
            B[2] = (1 - rate)*B[2] + rate*Bz
    if r[1] - ymin < boundarySize: # basic absorber
        rate = supRate*cos2shape(r[1] - ymin, 0, boundarySize)
        E[0] = (1 - rate)*E[0]
        E[1] = (1 - rate)*E[1]
        E[2] = (1 - rate)*E[2]
        B[0] = (1 - rate)*B[0]
        B[1] = (1 - rate)*B[1]
        B[2] = (1 - rate)*B[2]

#-----------------------------seting plasma-------------------------------------
N_cr = electron_mass * omega ** 2 / (4 * np.pi * electron_charge ** 2)
density = 100*N_cr
debye_length = 0.08165*wavelength/64.0
temperature = 4 * np.pi * density * (electron_charge ** 2) * debye_length ** 2
particlesPerCell = 100

@cfunc(types.add_particles_callback)
def density_callback(r, data_double, data_int):# callback function 
    return density*cos2shape(r[1] + wavelength, wavelength, wavelength) 
sim.add_particles(name='electron', number=int(nx*ny*0.25*particlesPerCell),
                 charge=-electron_charge, mass=electron_mass,
                 temperature=temperature, density=density_callback.address)


#=================================OUTPUT========================================
fig, axs = plt.subplots(2, constrained_layout=True)
ax = axs[0]

#------------------------------field output-------------------------------------
tmp = np.zeros((1, 1), dtype=np.double) # null plot to show the color bar
G_cmap=mpl.colors.LinearSegmentedColormap.from_list('N',[(1,1,1),(0,0.5,0)],N=32)
cbar = ax.imshow(tmp, vmin=0, vmax=2,
                 extent=[xmin,xmax,ymin,ymax], interpolation='none',
                 aspect='equal', origin='lower', cmap=G_cmap)
oBz = np.zeros((ny, nx), dtype=np.double)
maxField = 2*fieldAmplitude
plot0 = ax.imshow(oBz, vmin=-maxField, vmax=maxField,
                  extent=[xmin,xmax,ymin,ymax], interpolation='none',
                  aspect='equal', cmap='RdBu', origin='lower')
ax.set(xlabel='$x$ (cm)', ylabel='$y$ (cm)')
ax.ticklabel_format(axis='both', scilimits=(0,0), useMathText=True)
divider = make_axes_locatable(ax)
cax = divider.append_axes("right", size="2%", pad=0.05)
fig.add_axes(cax)
bar0 = fig.colorbar(plot0, cax=cax, orientation="vertical")
bar0.set_label(label='$B_z$ (CGS)')

@cfunc(types.field_loop_callback)
def fieldBz_cb(ind, r, E, B, data_double, data_int):
    Bz = carray(data_double, oBz.shape, dtype=np.double)
    Bz[ind[1], ind[0]] = B[2]

def plot_field():
    sim.field_loop(handler=fieldBz_cb.address, data_double=pipic.addressof(oBz))
    plot0.set_data(oBz)

#----------------------------density output-------------------------------------
oN = np.zeros((ny, nx), dtype=np.double)

@cfunc(types.particle_loop_callback)
def density_cb(r, p, w, id, data_double, data_int):   
    ix = int(oN.shape[1]*(r[0] - xmin)/(xmax - xmin))
    iy = int(oN.shape[0]*(r[1] - ymin)/(ymax - ymin))
    data = carray(data_double, oN.shape, dtype=np.double)
    if 0 <= iy < oN.shape[0] and 0 <= ix < oN.shape[1]:
        data[iy, ix] += w[0]/(dx*dy*2*density)

N_cmap=mpl.colors.LinearSegmentedColormap.from_list('N',[(0,0.5,0),(0,0.5,0)],N=9)
plot1 = ax.imshow(oN, vmin=0, vmax=1, alpha=oN,
                      extent=[xmin,xmax,ymin,ymax], interpolation='none',
                      aspect='equal', origin='lower', cmap=N_cmap)
ax.set(xlabel='$x$ (cm)', ylabel='$y$ (cm)')
ax.ticklabel_format(axis='both', scilimits=(0,0), useMathText=True)
cax1 = divider.append_axes("right", size="2%", pad=0.85)
fig.add_axes(cax1)
bar1 = fig.colorbar(cbar, cax=cax1, orientation="vertical")
bar1.set_label(label='$N/N_0$')

def plot_density():
    oN.fill(0)
    sim.particle_loop(name='electron', handler=density_cb.address,
                      data_double=pipic.addressof(oN))
    plot1.set_data(oN)

#----------------------------1dfield output-------------------------------------
oEp = np.zeros((2*nx,),dtype=np.double) #field P-component of the pulse
EpSize = 8*wavelength # size of the output

@cfunc(types.it2r_callback)
def E_it2r(it, r, data_double, data_int):
    lCoord = (it[0] + 0.5)*EpSize/oEp.shape[0]
    r[0] = lCoord*np.sin(incidenceAngle)
    r[1] = lCoord*np.cos(incidenceAngle)
    r[2] = 0

@cfunc(types.field2data_callback)
def get_Ep(it, r, E, B, data_double, data_int):
    data_double[it[0]]=-E[0]*np.cos(incidenceAngle)+E[1]*np.sin(incidenceAngle)

@cfunc(types.field2data_callback)
def get_Es(it, r, E, B, data_double, data_int):
    data_double[it[0]] = E[2]

axs[1].set_xlim([0, EpSize])
axs[1].set_ylim([-maxField, maxField])
axs[1].set(xlabel='$x$ (cm)', ylabel='$E_p$ (cgs units)')
x_axis = np.linspace(0, EpSize, oEp.shape[0])
plot_Ep, = axs[1].plot(x_axis, oEp)

with open('res_.npy', 'rb') as f: # reading data of RES computation
    res_X_0 = np.load(f)
    res_Ep = -np.load(f)
    res_Es = -np.load(f)

res_X = res_X_0 + -arrivalDelay * light_velocity
plot_res_Ep, = axs[1].plot(res_X, res_Ep)

def plot_E(i):
    sim.custom_field_loop(number_of_iterations=oEp.shape[0], it2r=E_it2r.address,
                          field2data=get_Ep.address, 
                          data_double=pipic.addressof(oEp))
    plot_Ep.set_ydata(oEp)
    res_X = res_X_0 + ((-arrivalDelay + i*time_step) * light_velocity +
                       wavelength * np.cos(incidenceAngle)) 
                       # this is due to surface being at 0.5 \mu m
    plot_res_Ep.set_xdata(res_X)
    if i == 21*figStride:
        with open('im21_Ep_' + str(nx) + '.npy', 'wb') as f:
            np.save(f, x_axis)
            np.save(f, oEp)


#===============================SIMULATION======================================
outputFolder = 'laser_solid_interaction_output'
if not os.path.exists(outputFolder):
   os.makedirs(outputFolder)
data_int = np.zeros((1, ), dtype=np.intc) # data for passing the iteration number
for i in range(figStride*21 + 1):
    print(i, '/', figStride*21 + 1)
    data_int[0] = i
    sim.field_loop(handler=field_callback.address, 
                   data_int=pipic.addressof(data_int), use_omp=True)
    if i%figStride == 0:
        plot_field()
        plot_density()
        plot_E(i)
        fig.savefig(outputFolder+'/im'+str(int(i/figStride))+'.png',dpi=300)
    sim.advance(time_step=time_step)
\end{minted}

\end{appendices}

\bibliography{literature}

\end{document}